\documentstyle{article}
\begin{document}
\title{
{\Large BRST} Invariance and Renormalisability of the SU(2)$\times$U(1)\\
          Electroweak Theory with Massive W Z Bosons}
\author{Ze-sen Yang, Xianhui Li, Zhinig Zhou and Yushu Zhong \\
Department of Physics,  Peking University, Beijing 100871, CHINA }
\date{\today}
\maketitle             
\begin{abstract}
     Since the SU(n) gauge theory with massive gauge bosons has been proven
to be renormalisable we reinvestigate the renormalisability of the
 SU$_L$(2) $\times$ U$_Y$(1) electroweak theory with massive W Z bosons. We
expound that with the constraint conditions caused by the W Z mass term and
the additional condition chosen by us we can  performed the quantization and
construct the ghost action in a way similar to that used for the massive
SU(n) theory. We also show
that when the $\delta-$ functions appearing in the path integral of the
Green functions and representing the constraint conditions are rewritten as
Fourier integrals with Lagrange multipliers $\lambda_a$ and $\lambda_y$,
the BRST invariance is kept in the total effective action consisting of the
Lagrange multipliers, ghost fields and the original fields. Furthermore, by
comparing with the massless theory and with the massive SU(n) theory we find
the general form of the divergent part of the generating functional for the
regular vertex functions and prove the renormalisability of the theory. It
is also clarified that the renormalisability of the theory with the  W Z
mass term is ensured by that of the massless theory and the massive SU(n)
theory.
\end{abstract}
\newpage
\begin{center}
{\bf I}.\ \ Introduction
\end{center} \par \ \par
     Although the negative answer to the problem of renormalisability  of
a SU(n) theory with massive gauge bosons is widely known, such theories 
continue to be studied (see for example Refs. [1-8]). However, since the
negative answer had not been voted down, it was naturally difficult to
investigate the possibility of directly adding a mass term to
the SU$_L$(2) $\times$ U$_Y$(1) theory. Recently, the renormalisability of
the massive SU(n) gauge theory has been proven [1,2]. Therefore we will
reinvestigate the SU$_L$(2) $\times$ U$_Y$(1) theory of S.L.Glashow [9]
with the mass term of the W Z fields. The study of the theory including the
mass term of the matter fields as well w1ll be reported in Ref. [10].
\par
    In order to make appropriate the mass ratio, the W Z mass term must
contain a product of the SU$_L$(2) and U$_Y$(1) fields and thus cause
constraint conditions containing products of such fields. Next, such a
mass term is invariant under an infinitesimal gauge transformation with
$\delta \theta_1$ and $\delta \theta_2$ equal to zero and $\delta \theta_3$
equal to $\delta \theta_y$, where $\theta_a$ and $\theta_1$ are the
parameters of the gauge group. Therefore an additional constraint condition
should be properly chosen. We will expound that with the constraint
conditions caused by the W Z mass term and the additional condition chosen
by us we can  performed the quantization and construct the ghost action in
a way similar to that used for the massive SU(n) theory [1]. We will also
show that when the $\delta-$ functions appearing in the path integral of the
Green functions and representing the constraint conditions are rewritten as
Fourier integrals with Lagrange multipliers $\lambda_a$ and $\lambda_y$,
the BRST invariance is kept in the total effective action consisting of the
Lagrange multipliers, ghost fields and the original fields.
\par
    As the constraint conditions contain the products of the SU$_L$(2) and
U$_Y$(1) fields, the divergent part of the generating functional $\Gamma$
for the regular vertex functions is dependent on the classical fields of
the Lagrange multipliers $\lambda_a$ and $\lambda_y$ when the generating
functional for the Green functions contains the sources of these Lagrange
multipliers. The problem of whether such a generalized form of the theory
is renormalisable becomes complicated. However, we are not interested in
using the Green functions involving $\lambda_a$ or $\lambda_y$. Thus we
can avoid introducing the sources of these Lagrange multipliers  to the
generating functional for the Green functions. An equivalent and convenient
procedure is to derive the Slavnov--Taylor identities and the additional
identities for $\Gamma$ 
with the help of the generalized form of the theory and then let vanish
the functional derivatives of $\Gamma$ with respect to the classical fields
of these Lagrange multipliers. In this way the divergent part of $\Gamma$
will be shown to satisfy the same equations appearing in the massless theory.
Furthermore, by comparing with the massless theory and with the massive
SU(n) theory we will be able to find the general form of the divergent part
of $\Gamma$ and prove the renormalisability of the theory. Meanwhile it will
be clarified that the renormalisability of the theory with the  W Z mass
term is ensured by that of the massless theory and the massive SU(n) theory.
\par
    In section $2$ we will find the constraint conditions caused by the
W Z mass term. The additional constraint condition will also be chosen.
The method of quantization will be explained in section $3$. Setion $4$ is
devoted to prove the renormalisability of the theory. Concluding remarks
will be given in the final section.
\par
\vspace{8mm}
\def\theequation{2.\arabic{equation}}
\begin{center}
{\bf II}.\ \ Original and Additional Constraint Conditions
\end{center} \par \ \par
    For the sake of comvenience we assume in the present work that the
matter fields consist only of the electron and electron-neutrino fields
and are often denoted by $\psi(x)$ and $\overline{\psi}(x)$. The former
stands for the purely left-handed neutrino field $\nu_L$, the left- and
right-handed parts of the electron field namely $e_L$, $e_R$, and the
latter stands for $\overline{\nu}_L$, $\overline{e}_L$ and $\overline{e}_R$.
Next let $W_{a \mu}(x)$, $W_{y \mu}(x)$ be the SU$_L$(2) and U$_Y$(1) gauge
fields and $g$, $g_1$ be the coupling constants. Thus the W Z mass term in
the Lagrangian is
\begin{eqnarray}
{\cal L}_{WM} = \frac{1}{2} M^2 W_{a \mu}W_{a}^{\mu} 
          + \frac{1}{2} M^2 \Big(\frac{g_1}{g}\Big)^2 W_{y \mu}W_{y}^{\mu}
          - M^2 \Big(\frac{g_1}{g}\Big) W_{3 \mu}W_{y}^{\mu} \,,
\end{eqnarray}
or
$$
{\cal L}_{WM} = \frac{1}{2} M^2 W_{1 \mu}(x) W_{1}^{\mu}(x)
                 + \frac{1}{2} M^2 W_{2 \mu}(x) W_{2}^{\mu}(x)
                 + \frac{1}{2} M_z^2 Z_{\mu}(x) Z^{\mu}(x) \,, 
$$ 
where $M_z^2$ stands for $g^{-2}( g^2+ g_1^2) M^2$, and $Z_{\mu}(x)$,
$A_{\mu}(x)$ are the field functions of Z boson and photon, namely 
\begin{eqnarray} 
&& Z_{\mu} =\frac{1}{\sqrt{( g^2+ g_1^2)}} 
  (gW_{3 \mu} - g_1 W_{y \mu})\,, \\
&& A_{\mu} =\frac{1}{\sqrt{( g^2+ g_1^2)}}
  \varepsilon (g_1 W_{3 \mu} + g W_{y \mu}) \,,
\end{eqnarray}
where $\varepsilon $ is $1$ or $-1$.
\par
    The original Lagrangian of the SU$_L$(2) $\times$ U$_Y$(1) electroweak
theory with the mass term ${\cal L}_{WM}$ is
\begin{eqnarray}
&{\cal L} =  {\cal L}_{\psi} + {\cal L}_{\psi W} + {\cal L}_{WM}
              + {\cal L}_{WL} + {\cal L}_{WY} \,,
\end{eqnarray}
where ${\cal L}_{\psi}$ describe the pure matter fields,
${\cal L}_{\psi W}$ is the coupling term between the matter and gauge
fields and 
\begin{eqnarray}
&&{\cal L}_{WL} = -\frac{1}{4} F_{a \mu \nu} F_a^{\mu \nu} \,, \\
&&{\cal L}_{WY} = -\frac{1}{4} B_{\mu \nu} B^{\mu \nu} \,,
\end{eqnarray}
with              
\begin{eqnarray}
&& F_{a \mu \nu} = \partial_{\mu}W_{a \nu}-\partial_{\nu}W_{a \mu}
- g C_{abc} W_{b \mu } W_{c \nu} \,, \\
&& B_{\mu \nu} = \partial_{\mu}W_{y \nu}-\partial_{\nu}W_{y \mu}\,.
\end{eqnarray}
$C_{abc}$ stands for the structure constants of SU$_L$(2) with $C_{123}$
equal to $1$.
\par
    Denote by $ \theta_a(x), \theta_y(x) $ the parameters of the gauge group.
Thus, under an infinitesimal gauge transformation, the fields $W_a^{\mu}$,
$W_y^{\mu}$, $\psi$ and $\overline{\psi}$ transform as
\begin{eqnarray*}
&& \delta W_a^{\mu}(x) =
    - \frac{1}{g} \partial^{\mu} \delta \theta_a(x)
     - C_{abc} W_c^{\mu}(x) \delta \theta_b(x)  \,,  \\
&& \delta W_y^{\mu}(x) =
    - \frac{1}{g_1} \partial^{\mu} \delta \theta_y(x) \,, \\
&& \delta \nu_L(x) = \frac{i}{2}\delta \theta_1(x) e_L(x)
+ \frac{1}{2}\delta \theta_2(x)e_L(x)+ \frac{i}{2}\delta \theta_3(x)\nu_L(x)
   - \frac{i}{2}\delta \theta_y(x) \nu_L(x)\,, \\
&& \delta e_L(x) = \frac{i}{2}\delta \theta_1(x) \nu_L(x)
 - \frac{1}{2}\delta \theta_2(x)\nu_L(x) -\frac{i}{2}\delta \theta_3(x)e_L(x)
 - \frac{i}{2}\delta \theta_y(x)e_L(x) \,,\\
&& \delta e_R(x) = -i \delta \theta_y(x) e_R(x) \,, \\
&&\delta \overline{\nu}_L(x) = -\frac{i}{2}\delta\theta_1(x)\overline{e}_L(x)
    + \frac{1}{2}\delta \theta_2(x) \overline{e}_L(x)
    - \frac{i}{2}\delta \theta_3(x) \overline{\nu}_L(x)
    + \frac{i}{2}\delta \theta_y(x) \overline{\nu}_L(x)\,, \\
&&\delta \overline{e}_L(x) = -\frac{i}{2}\delta\theta_1(x)\overline{\nu}_L(x)
    - \frac{1}{2}\delta \theta_2(x)\overline{\nu}_L(x)
    + \frac{i}{2} \delta \theta_3(x) \overline{e}_L(x)
    + \frac{i}{2}\delta \theta_y(x) \overline{e}_L(x) \,, \\
&&\delta \overline{e}_R(x) = i \delta \theta_y(x) \overline{e}_R(x)\,.
\end{eqnarray*}
Therefore the action transforms as
\begin{eqnarray}
&& \delta \int d^4x {\cal L}(x) =
  \delta \int d^4x {\cal L}_{WM}(x) \nonumber \\
&&\ \ \ \ \ \ \ \ =\,\int d^4x\Big\{
 \Big( \frac{M^2}{g} \partial_{\mu} W_1^{\mu}(x)
       + \frac{M^2}{g}g_1 W_{2 \mu}(x) W_y^{\mu}(x)  
       \Big) \delta \theta_1  \nonumber \\
&&\ \ \ \ \ \ \ \ \ \ \ \
 + \Big( \frac{M^2}{g} \partial_{\mu} W_2^{\mu}(x)
        - \frac{M^2}{g}g_1 W_{1 \mu}(x) W_y^{\mu}(x)  
        \Big) \delta \theta_2  \nonumber \\
&&\ \ \ \ \ \ \ \ \ \ \ \
 + \Big( \frac{M^2}{g} \partial_{\mu} W_3^{\mu}(x)
         - \frac{M^2}{g^2} g_1 \partial_{\mu}W_y^{\mu}(x)  
     \Big) ( \delta \theta_3 - \delta \theta_y )   \Big\} \,.
\end{eqnarray}
Since the classical equations of motion make the action invariant under
an arbitrary infinitesimal transformation of the field functions, they 
certainly make the W Z mass term invariant under an arbitrary infinitesimal
gauge transformation. This means that when $M$ is not equal to zero, the
classical equations of motion leads to the following constraint conditions 
\begin{eqnarray}
&& \frac{M^2}{g} \partial_{\mu} W_1^{\mu}(x)
      + \frac{M^2}{g}g_1 W_{2 \mu}(x) W_y^{\mu}(x) = 0 \,,\\
&& \frac{M^2}{g} \partial_{\mu} W_2^{\mu}(x)
      - \frac{M^2}{g}g_1 W_{1 \mu}(x) W_y^{\mu}(x) = 0 \,,\\
&& \frac{M^2}{g} \partial_{\mu} W_3^{\mu}(x)
       - \frac{M^2}{g^2} g_1 \partial_{\mu}W_y^{\mu}(x) = 0 \,.
\end{eqnarray}
These are the original constraint conditions. As it can be seen from (2.9)
that the W Z mass term is invariant under an infinitesimal gauge
transformation with $\delta \theta_1$ and $\delta \theta_2$ equal to zero
and $\delta \theta_3$ equal to $\delta \theta_y$. For this reason,  
$\partial_{\mu}W_3^{\mu} $ and $\partial_{\mu}W_y^{\mu} $ appear in one
constraint. We now choose an additional condition and replace (2.12) with
\begin{eqnarray}
&& \frac{M^2}{g} \partial_{\mu} W_3^{\mu}(x)
              + \frac{M^2}{g} g_1 W_{3 \mu}(x)W_y^{\mu}(x) = 0 \,,\\
&& \partial_{\mu}W_y^{\mu}(x) + g W_{3 \mu}(x) W_y^{\mu}(x) = 0 \,.
\end{eqnarray}
\par
\vspace{8mm}
\def\theequation{3.\arabic{equation}}
\begin{center}
{\bf III}.\ \ Quantization and BRST Invariance
\end{center} \par \ \par
\setcounter{equation}{0}
    Write (2.10), (2.11) and (2.13),(2.14) as
\begin{eqnarray}
 \Phi_a(x) = 0 \,,\ \ \ \ \ \  \Phi_y(x) = 0 \,,
\end{eqnarray}
with
\begin{eqnarray}
&& \Phi_1(x) = \partial_{\mu} W_1^{\mu}(x)
             + g_1 W_{2 \mu}(x) W_y^{\mu}(x)   \,,\\
&& \Phi_2(x) = \partial_{\mu} W_2^{\mu}(x)
             - g_1 W_{1 \mu}(x) W_y^{\mu}(x)   \,,\\
&& \Phi_3(x) = \partial_{\mu} W_3^{\mu}(x)
             + g_1 W_{3 \mu}(x)W_y^{\mu}(x)  \,, \\
&& \Phi_y(x) = \partial_{\mu}W_y^{\mu}(x)
             + g W_{3 \mu}(x) W_y^{\mu}(x)  \,.
\end{eqnarray}
\par
    Taking the constraint conditions (3.1) into account one should write the
path integral of the Green functions inolving only the original fields as
\begin{eqnarray}
 \frac{1}{N_0} \int {\cal D}[{\cal W},\overline{\psi},\psi] 
 \Delta [{\cal W},\overline{\psi},\psi]
\prod_{a',x'} \delta \left(\Phi_{a'}(x')\right)\delta\left(\Phi_y(x') \right)
W_{a \mu} (x)W_{b \nu}(y) \cdots {\rm exp} \{{\rm i} I \} \,,
\end{eqnarray} 
where
\begin{eqnarray*} 
&& I = \int d^4x {\cal L}(x) \,, \\
&& N_0 =  \int {\cal D}[{\cal W},\overline{\psi},\psi]
 \Delta [{\cal W},\overline{\psi},\psi]
\prod_{a',x'} \delta \left(\Phi_{a'}(x')\right)\delta\left(\Phi_y(x') \right) 
 {\rm exp} \{{\rm i} I \} \,.
\end{eqnarray*}
The weight factor $\Delta[{\cal W},\overline{\psi},\psi]$ is to be
determined. Since only the field functions which satisfy the constraint
conditions can play roles in the integral (3.6), the value of the Lagrangian
can be changed for the field functions which do not satisfy these conditions.
In view of the fact that the conditions (3.1) make the action invariant with
respect to the infinitesimal gauge trasformation, we now imagine to replace
the mass term ${\cal L}_{WM}$ in (3.6) with a gauge invariant mass term
which is equal to ${\cal L}_{WM}$ when the conditions (3.1) are satisfied.
Thus, analogous to the case in the Fadeev--Popov method [1,11-16],
$\Delta[{\cal W},\overline{\psi},\psi]$ should be gauge invariant and make
the following equation valid for an arbitrary gauge invariant quantity
${\cal O}({\cal W,\,\overline{\psi},\,\psi })$
\begin{eqnarray*}
&&  \int {\cal D}[{\cal W},\,\overline{\psi},\,\psi] 
 \Delta[{\cal W},\overline{\psi},\psi]
 \prod_{a',x'}
 \delta \left(\Phi_{a'}(x') \right)\delta\left(\Phi_y(x') \right)
 {\cal O}({\cal W},\overline{\psi},\psi)
{\rm exp} \{{\rm i} \widetilde{I} \} \\
&&\ \ \ \ \ \ \ \ \ \ \ \
 \propto  \int {\cal D}[{\cal W},\,\overline{\psi},\,\psi]
{\cal O}({\cal W},\overline{\psi},\psi)
{\rm exp} \{{\rm i} \widetilde{I} \}   \,.
\end{eqnarray*}
where $\widetilde{I}$ is a gauge invariant action constructed by replacing
${\cal L}_{WM}$ with the imagined mass term. This means that the weight
factor $\Delta[{\cal W},\overline{\psi},\psi]$ can be determined according
to the Fadeev--Popov equation of the following form 
\begin{eqnarray}
 \Delta[{\cal W},\overline{\psi},\psi]
\int \prod_z d \Omega (z) \prod_{\sigma,x} \delta
\Big( \Phi_{\sigma}^{\Omega}(x) \Big) = 1 \,.
\end{eqnarray}
where $\sigma$ stands for $1,2,3,y$, $\Phi_{\sigma}^{\Omega}(x)$ is the
result of acting on $\Phi_{\sigma}(x)$ with a gauge transformation having
the parameters of the element $\Omega(x)$ of the gauge group, $d\Omega (z)$
is the volume element of the group integral. It follows that with the F--P
ghost fields $C_a(x)$, $C_y(x)$, $\overline{C}_a(x)$, $\overline{C}_y(x)$
as new variables, one can express the ghost Lagrangian as
\begin{eqnarray}
 {\cal L}^{(C)}(x) = \overline{C}_a(x) \Delta \Phi_a(x)
                      + \overline{C}_y(x) \Delta \Phi_y(x)\,,
\end{eqnarray}
where $\Delta \Phi_a(x)$, $\Delta \Phi_y(x)$ are defined by the BRST
transformtion of $\Phi_a(x)$ and $\Phi_y(x)$ so that
\begin{eqnarray}
 \delta_B \Phi_a(x) = \delta \zeta \Delta \Phi_a(x) \,,\ \ \ \ 
  \delta_B \Phi_y(x) = \delta \zeta \Delta \Phi_y(x) \,,                     
\end{eqnarray}
where $\delta \zeta $ is an infinitesimal fermionic parameter independent
of $x$. The BRST transformation of the gauge fields or matter fields is
nothing but the infinitesimal gauge transformation with $\delta \theta_a$
and $\delta \theta_y$ equal to $-g \delta \zeta C_a $
and $-g_1 \delta \zeta C_y $ respectively. Namely
\begin{eqnarray}
&& \delta_B W_a^{\mu}(x)  = \delta \zeta \Delta W_a^{\mu}(x) 
                       = \delta \zeta D^{\mu}_{ab}C_b(x) \,,\\
&& \delta_B W_y^{\mu}(x) = \delta \zeta \Delta W_y^{\mu}(x) 
                      = \delta \zeta \partial^{\mu} C_y(x) \,,\\
&& \delta_B \psi(x) = \delta \zeta \Delta \psi(x) \,, \ \ \ \ \ 
  \delta_B \overline{\psi}(x) = \delta \zeta \Delta \overline{\psi}(x) \,,
\end{eqnarray}
where
\begin{eqnarray*}
&& D^{\mu}_{ab}(x) =  \delta_{ab} \partial^{\mu}
                     + g f_{abc} A_c^{\mu}(x)  \,, \\
&& \Delta \nu_L(x) = - \frac{i}{2} gC_1(x) e_L(x)
  - \frac{1}{2} gC_2(x)e_L(x) - \frac{i}{2} gC_3(x)\nu_L(x)
   + \frac{i}{2} g_1 C_y(x) \nu_L(x)\,,\\
&& \Delta e_L(x) = - \frac{i}{2} gC_1(x) \nu_L(x)
   + \frac{1}{2} gC_2(x)\nu_L(x) + \frac{i}{2} gC_3(x)e_L(x)
   + \frac{i}{2} g_1 C_y(x)e_L(x) \,,\\
&& \Delta e_R(x) = i g_1 C_y(x) e_R(x) \,,\\
&&\Delta \overline{\nu}_L(x) = \frac{i}{2} gC_1(x)\overline{e}_L(x)
    - \frac{1}{2} gC_2(x) \overline{e}_L(x)
    + \frac{i}{2} gC_3(x) \overline{\nu}_L(x)
    - \frac{i}{2} g_1 C_y(x) \overline{\nu}_L(x)\,,\\
&&\Delta \overline{e}_L(x) = \frac{i}{2} gC_1(x)\overline{\nu}_L(x)
    + \frac{1}{2} gC_2(x)\overline{\nu}_L(x)
    - \frac{i}{2} gC_3(x) \overline{e}_L(x)
    - \frac{i}{2} g_1 C_y(x) \overline{e}_L(x) \,,\\
&&\Delta \overline{e}_R(x) = -i g_1 C_y(x) \overline{e}_R(x)\,.
\end{eqnarray*}
 $ C_a(x)$ and $C_y(x) $ are also transformed as usual
\begin{eqnarray*}
&& \delta_B C_a(x) = \delta \zeta \Delta C_a(x) 
                = \delta \zeta \frac{g}{2} C_{abc}C_b(x)C_c(x) \,,\\
&& \delta_B C_y(x) = 0 \,.
\end{eqnarray*}
Now we can write $\Delta \Phi_a(x)$, $ \Delta \Phi_y(x)$ as
\begin{eqnarray}
&&  \Delta \Phi_1 = \partial_{\mu}\Delta W_1^{\mu}(x)
                  + g_1 \Delta W_2^{\mu}(x)  W_{y \mu}(x)
                  + g_1 W_{2 \mu}(x)\Delta W_y^{\mu}(x) \,, \\
&&  \Delta \Phi_2 =  \partial_{\mu}\Delta W_2^{\mu}(x)
                  - g_1 \Delta W_1^{\mu}(x)  W_{y \mu}(x)
                  - g_1 W_{1 \mu}(x)\Delta W_y^{\mu}(x) \,, \\
&&  \Delta \Phi_3 = \partial_{\mu}\Delta W_3^{\mu}(x)
                  + g_1 \Delta W_3^{\mu}(x)  W_{y \mu}(x)
                  + g_1 W_{3 \mu}(x)\Delta W_y^{\mu}(x) \,, \\
&&  \Delta \Phi_y = \partial_{\mu}\Delta W_y^{\mu}(x)
                  + g \Delta W_3^{\mu}(x) W_{y \mu}(x)
                  + g W_{3 \mu}(x)\Delta W_y^{\mu}(x) \,,  
\end{eqnarray}
Since $\Delta W_a^{\mu}$, $\Delta W_y^{\mu}$, $\Delta\psi(x)$,
 $\Delta\overline{\psi}(x)$ and $\Delta C_a(x)$ are BRST invariant, it is
easy to see that $\Delta \Phi_a(x)$ and $\Delta \Phi_y(x)$ are also BRST
invariant.
\par
    One can further generalized the theory by regarding as new variables
the Lagrange multipliers $\lambda_a(x)$ and $\lambda_y(x)$ associated with
the constraint conditions. Thus the total effective Lagrangian and action
consist of these Lagrange multipliers, ghosts and the original variables, 
namely
\begin{eqnarray}
&&{\cal L}_{{\rm eff}}(x) = {\cal L}(x) + {\cal L}^{(C)}(x)
                       + \lambda_a(x) \Phi_a(x) 
                       + \lambda_y(x) \Phi_y(x)  \,, \\
&& I_{{\rm eff}} = \int d^4x {\cal L}_{{\rm eff}}(x)  \,. 
\end{eqnarray}
Correspondingly, the path integral of the generating functional for the
Green functions is
\begin{eqnarray}
&& {\cal Z}[\overline{\eta}, \eta,\overline{\chi},\chi,J,j]
= \frac{1}{N_{\lambda}}
\int {\cal D}[\overline{\psi},\psi,{\cal W}, \overline{C},C,\lambda]
 {\rm exp} \Big\{ {\rm i} \big( I_{{\rm eff}} + I_s \big) \Big\},
\end{eqnarray}
where $N_{\lambda}$ is a constant, $I_s$ is the source term in the action.
They are defined by
\begin{eqnarray}  
&& N_{\lambda}
   = \int {\cal D}[\overline{\psi}, \psi,{\cal W},\overline{C},C,\lambda]
     {\rm exp} \Big\{ {\rm i} I_{{\rm eff}}  \Big\} \,, \nonumber \\
&&I_s = \int d^4x \Big\{\overline{\eta}(x)\psi(x) + \overline{\psi}(x)\eta(x)  
          + \overline{\chi}_a(x)C_a(x) + \overline{C}_a(x) \chi_a(x)  
          + \overline{\chi}_y(x)C_y(x) \nonumber\\
&&\ \ \ \ \ \ \ \ 
          + \overline{C}_y(x) \chi_y(x)
          + J_a^{\mu}(x) W_{a \mu}(x) + J_y^{\mu}(x) W_{y \mu}(x)
          + j_a(x) \lambda_a(x) + J_y(x) \lambda_y(x)  \Big\} \,,
\end{eqnarray}
where $\overline{\eta}(x),\eta(x)\cdots$ stand for the sources. In particular,
 $j_a(x)$, $j_y(x)$ are the sources of $\lambda_a(x)$, $\lambda_y(x)$,
 respectively. 
\par
    We now check the BRST invariance of the effective action $I_{eff}$
defined by (3.17) and (3.18). With ${\overline C}_a(x)$, ${\overline C}_y(x)$
transforming as 
$$
 \delta_B {\overline C}_a(x) = - \delta \zeta \lambda_a(x)\,, \ \ \ \ 
 \delta_B {\overline C}_y(x) = - \delta \zeta \lambda_y(x)\,,
$$
and noticing the invariance of $\Delta\Phi_a, \Delta\Phi_y$, one has
$$
 \delta_B \int d^4x {\cal L}^{(C)}(x)
  = \int d^4x \Big\{ -\lambda_a(x) \delta_B \Phi_a(x)
                    - \lambda_y(x) \delta_B \Phi_y(x) \Big\} \,.
$$
Therefore
$$
 \delta_B I_{eff} 
  = \delta_B I_{WM} 
         + \int d^4x \Big\{ \big( \delta_B\lambda_a(x) \big) \Phi_a(x)
                   + \big( \delta_B \lambda_y(x) \big) \Phi_y(x) \Big\} \,.
$$
From this and the expression of $\delta_B I_{WM}$, it can be shown that
the effective action is invariant, when the transformation of $\lambda_a(x)$
and $\lambda_y(x)$ are defined as
\begin{eqnarray*}
&& \delta_B \lambda_1(x) = \delta \zeta M^2 C_1(x) \,,\\
&& \delta_B \lambda_2(x) = \delta \zeta M^2 C_2(x) \,,\\
&& \delta_B \lambda_3(x)
    = \delta \zeta M^2 C_3(x)
       - \delta \zeta \frac{g_1}{g} M^2 C_y(x)  \,,\\
&& \delta_B \lambda_y(x)
= \delta \zeta \frac{g_1^2}{g^2} M^2 C_y(x)
  - \delta \zeta \frac{g_1}{g} M^2 C_3(x)  \,.
\end{eqnarray*}
\par
\vspace{8mm}
\def\theequation{4.\arabic{equation}}
\begin{center}
{\bf IV}.\ \ Renormalisability 
\end{center} \par \ \par
\setcounter{equation}{0}
    Let $W_{a \mu}(x), W_{y \mu}(x) $, $C_a(x), C_y(x), \cdots $ stand
for the renormalized field founctions, $g, g_1$ and $M$ be renormalized
parameters. By introducing the source terms of the composite field
functions  $\Delta W_a^{\mu}$, $\Delta W_y^{\mu}$, $\Delta C_a(x)$,
$\Delta \psi(x)$, $\Delta \overline{\psi}(x)$ and the sources $K^a_{\mu}(x)$,
$K^y_{\mu}(x)$, $L_a(x)$, $n_{\alpha}(x)$, $l_{\alpha}(x)$, $p_{\alpha}(x)$,
 $n'_{\alpha}(x)$, $l'_{\alpha}(x)$ and $p'_{\alpha}(x)$, the effective
Lagrangian without counterterm becomes
\begin{eqnarray}
 {\cal L}^{[0]}_{eff}(x)  
    &=& \lambda_a(x) \Phi_a(x)
       + \lambda_y(x) \Phi_y(x) 
       + {\cal L}_{WL}(x) + {\cal L}_{WY}(x) \nonumber \\
    &&+ {\cal L}_{WM}(x) + {\cal L}^{(C)}(x) 
      + {\cal L}_{\psi}(x) + {\cal L}_{\psi W}(x) \nonumber \\
    &&+ K^a_{\mu}(x) \Delta W_a^{\mu}(x)
      + K^y_{\mu}(x) \Delta W_y^{\mu}(x)
      + L_a(x) \Delta C_a(x) \nonumber\\ 
    &&+ n_{\alpha}(x) \Delta \nu_{L \alpha}(x)
      + l_{\alpha}(x) \Delta e_{L \alpha}(x)
      + p_{\alpha}(x) \Delta e_{R \alpha}(x) \nonumber\\
    &&+ n'_{\alpha}(x) \Delta \overline{\nu}_{L \alpha}(x)
      + l'_{\alpha}(x) \Delta \overline{e}_{L \alpha}(x)
      + p'_{\alpha}(x) \Delta \overline{e}_{R \alpha}(x)   \,.
\end{eqnarray}
The complete effective Lagrangian is the sum of ${\cal L}^{[0]}_{eff}$ and
the counterterm ${\cal L}_{count}$ 
\begin{eqnarray}
 {\cal L}_{{\rm eff}} = {\cal L}^{[0]}_{{\rm eff}} + {\cal L}_{count} \,.
\end{eqnarray}
\par
With (4.1), the generating functional for Green functions is defined as
\begin{eqnarray}
{\cal Z}^{[0]}
     [\overline{\eta}, \eta,\overline{\chi},\chi,J,j,K,L,n,l,p,n',l',p']
 = \frac{1}{N} \int {\cal D}
      [\overline{\psi},\psi,{\cal W}, \overline{C},C,\lambda]
   {\rm exp} \Big\{{\rm i} \big( I^{[0]}_{eff} + I_s \big) \Big\}\,, 
\end{eqnarray}
 $I^{[0]}_{eff}$ is the effective action $\int d^4x{\cal L}^{[0]}_{eff}(x)$,
$N$ is a constant to make ${\cal Z}^{[0]}$ equal to $1$ in the absence of
\begin{eqnarray*}
&&I_s = \int d^4x \Big\{\overline{\eta}(x)\psi(x) + \overline{\psi}(x)\eta(x)
          + \overline{\chi}_a(x)C_a(x) + \overline{C}_a(x) \chi_a(x)  
          + \overline{\chi}_y(x)C_y(x) \\
&&\ \ \ \ \ \ \ \ 
          + \overline{C}_y(x) \chi_y(x)
          + J_a^{\mu}(x) W_{a \mu}(x) + J_y^{\mu}(x) W_{y \mu}(x)
          + j_a(x) \lambda_a(x) + j_y(x) \lambda_y(x)  \Big\} \,,
\end{eqnarray*}
where $\overline{\eta}\, \psi$ and $\overline{\psi}\, \eta$ stand for
\begin{eqnarray*}  
&&\overline{\eta}\, \psi = 
        \overline{\eta}^{(\nu)}_{\alpha}\, \nu_{L \alpha}
      + \overline{\eta}^{(l)}_{\alpha}\,  e_{L \alpha}
      + \overline{\eta}^{(r)}_{\alpha}\,  e_{R \alpha} \,,\\
&&\overline{\psi}\, \eta =
        \overline{\nu}_{L \alpha}\, \eta^{(\nu)}_{\alpha}
      + \overline{e}_{L \alpha}\, \eta^{(l)}_{\alpha}
      + \overline{e}_{R \alpha}\, \eta^{(r)}_{\alpha}  \,.
\end{eqnarray*}
Denoting by ${\cal W}^{[0]}$ and $\Gamma^{[0]}$ the generating functionals
for connected Green functions and regular vertex functions respectively,
one has
\begin{eqnarray}
&&{\cal Z}^{[0]} 
= {\rm exp} \Big\{ {\rm i}{\cal W}^{[0]}
 [\overline{\eta},\eta,\overline{\chi},\chi,J,j,K,L,n,l,p,n',l',p'] \Big\}\,,
\\
&& \Gamma^{[0]}
 [\widetilde{\psi},\widetilde{\overline{\psi}},\widetilde{W},
  \widetilde{\overline{C}}, \widetilde{C},
  \widetilde{\lambda},K,L,n,l,p,n',l',p'] \nonumber\\
&& \ \ \ \ \ = {\cal W}^{[0]} 
        - \int d^4x \Big[ J_a^{\mu} \widetilde{W}_{a \mu}
       + J_y^{\mu} \widetilde{W}_{y \mu}
       + j_a \widetilde{\lambda}_a
       + j_y \widetilde{\lambda}_y 
       + \overline{\chi}_a \widetilde{C}_a
       + \widetilde{\overline{C}}_a \chi_a
       + \overline{\chi}_y \widetilde{C}_y  \nonumber\\
&& \ \ \ \ \ \ \ \ \ \ \ \ \ \ \ \ \ \ \
       + \widetilde{\overline{C}}_y \chi_y
       + \overline{\eta}^{(\nu)} \widetilde{\nu}_L
       + \overline{\eta}^{(l)} \widetilde{e}_L 
       + \overline{\eta}^{(r)} \widetilde{e}_R 
       + \widetilde{\overline{\nu}}_L \eta^{(\nu)}
       + \widetilde{\overline{e}}_L \eta^{(l)}
       + \widetilde{\overline{e}}_R \eta^{(r)}   \Big]\,,
\end{eqnarray}
where $\widetilde{W}_{a \mu}$, $\widetilde{\nu}_L$, $\cdots$ are the
so-called classical fields defined by 
\begin{eqnarray*}
&& \widetilde{W}_{a \mu}(x) = \frac{\delta {\cal W}^{[0]}}
                                  {\delta J_a^{\mu}(x) }\,,\ \ \ \ \ \ 
 \widetilde{\lambda}_a(x) = \frac{\delta {\cal W}^{[0]}}
                                  {\delta j_a(x) }\,,\ \ \ \ \ \ 
 \widetilde{C}_a(x) = \frac{\delta {\cal W}^{[0]}}
                                  {\delta \overline{\chi}_a(x)}\,,
\nonumber\\
&& \widetilde{\overline{C}}_a(x) = - \frac{\delta {\cal W}^{[0]}}
                                         {\delta \chi_a(x) }\,,\ \ \ \ \ \
 \widetilde{W}_{y \mu}(x) = \frac{\delta {\cal W}^{[0]}}
                                  {\delta J_y^{\mu}(x) }\,,\ \ \ \ \ \ 
 \widetilde{\lambda}_y = \frac{\delta {\cal W}^{[0]}}
                                  {\delta j_y(x) }\,, 
\nonumber\\
&& \widetilde{C}_y(x) = \frac{\delta {\cal W}^{[0]}}
                                  {\delta \overline{\chi}_y(x)}\,,\ \ \ \ \ \
 \widetilde{\overline{C}}_y(x) = - \frac{\delta {\cal W}^{[0]}}
                                         {\delta \chi_y(x) }\,,\ \ \ \ \ \
 \widetilde{\nu}_{L \alpha}(x) = \frac{\delta {\cal W}^{[0]}}
              {\delta \overline{\eta}^{(\nu)}_{\alpha}(x) }\,, 
\nonumber\\
&& \widetilde{e}_{L \alpha}(x) = \frac{\delta {\cal W}^{[0]}}
               {\delta \overline{\eta}^{(l)}_{\alpha}(x)}\,,\ \ \ \ \ \
 \widetilde{e}_{R \alpha}(x) = \frac{\delta {\cal W}^{[0]}}
               {\delta \overline{\eta}^{(r)}_{\alpha}(x) }\,,\ \ \ \ \ \
 \widetilde{\overline{\nu}}_{L \alpha}(x) = - \frac{\delta {\cal W}^{[0]}}
                           {\delta \eta^{(\nu)}_{\alpha}(x) }\,, 
\nonumber\\
&& \widetilde{\overline{e}}_{L \alpha}(x) = - \frac{\delta {\cal W}^{[0]}}
                           {\delta \eta^{(l)}_{\alpha}(x)}\,,\ \ \ \ \ \
 \widetilde{\overline{e}}_{R \alpha}(x) = - \frac{\delta {\cal W}^{[0]}}
               {\delta \eta^{(r)}_{\alpha}(x) }\,.
\end{eqnarray*}
Therefore 
\begin{eqnarray*}
&& J_a^{\mu}(x) = - \frac{\delta \Gamma^{[0]}}
                        {\delta \widetilde{W}_{a \mu}(x) } \,, \ \ \ \ \ \ 
 j_a(x) = - \frac{\delta \Gamma^{[0]}}
                        {\delta \widetilde{\lambda}_a(x) } \,, \ \ \ \ \ \ 
 \overline{\chi}_a(x) = \frac{\delta \Gamma^{[0]}}
                              {\delta \widetilde{C}_a(x)} \,, \\ 
&& \chi_a(x) = - \frac{\delta \Gamma^{[0]}}
                     {\delta \widetilde{\overline{C}}_a(x) }\,, \ \ \ \ \ \ 
 J_y^{\mu}(x) = - \frac{\delta \Gamma^{[0]}}
                        {\delta \widetilde{W}_{y \mu}(x) } \,, \ \ \ \ \ \ 
 j_y(x) = - \frac{\delta \Gamma^{[0]}}
                        {\delta \widetilde{\lambda}_y(x) } \,,  \\
&& \overline{\chi}_y(x) = \frac{\delta \Gamma^{[0]}}
                              {\delta \widetilde{C}_y(x)} \,, \ \ \ \ \ \ 
 \chi_y(x) = - \frac{\delta \Gamma^{[0]}}
                     {\delta \widetilde{\overline{C}}_y(x) }\,, \ \ \ \ \ \ 
 {\eta}^{(\nu)}_{\alpha}(x) = - \frac{\delta \Gamma^{[0]}}
              {\delta \widetilde{\overline{\nu}}_{L \alpha}(x) }\,,\\ 
&&{\eta}^{(l)}_{\alpha}(x)  = - \frac{\delta \Gamma^{[0]}}
               {\delta \widetilde{\overline{e}}_{L \alpha}(x)}\,,\ \ \ \ \ \
{\eta}^{(r)}_{\alpha}(x)  = - \frac{\delta \Gamma^{[0]}}
              {\delta \widetilde{\overline{e}}_{R \alpha}(x) }\,, \ \ \ \ \ \
 \overline{\eta}^{(\nu)}_{\alpha}(x) =  \frac{\delta \Gamma^{[0]}}
              {\delta \widetilde{\nu}_{L \alpha}(x) }\,, \\
&&\overline{\eta}^{(l)}_{\alpha}(x)  =  \frac{\delta \Gamma^{[0]}}
               {\delta \widetilde{e}_{L \alpha}(x)}\,,\ \ \ \ \ \
\overline{\eta}^{(r)}_{\alpha}(x)  =  \frac{\delta \Gamma^{[0]}}
               {\delta \widetilde{e}_{R \alpha}(x) }\,.
\end{eqnarray*}
Besides, for $K^a_{\mu}, L_a$ $\cdots$, the spectators in the Legendre
transtrormation, one has
\begin{eqnarray*}
&& \frac{\delta {\cal W}^{[0]}}{\delta K^a_{\mu}(x) }
       = \frac{\delta \Gamma^{[0]}}{\delta K^a_{\mu}(x) } \,, \ \ \ \ \ \
 \frac{\delta {\cal W}^{[0]}}{\delta K^y_{\mu}(x) }
       = \frac{\delta \Gamma^{[0]}}{\delta K^y_{\mu}(x) } \,, \ \ \ \ \ \
 \frac{\delta {\cal W}^{[0]}}{\delta L_a(x)}
       = \frac{\delta \Gamma^{[0]}}{\delta L_a(x)} \,, \ \ \ \ \ \
\\
&& \frac{\delta {\cal W}^{[0]}}{\delta n_{\alpha}(x) }
       = \frac{\delta \Gamma^{[0]}}{\delta n_{\alpha}(x) } \,, \ \ \ \ \ \
 \frac{\delta {\cal W}^{[0]}}{\delta l_{\alpha}(x) }
       = \frac{\delta \Gamma^{[0]}}{\delta l_{\alpha}(x) } \,, \ \ \ \ \ \
 \frac{\delta {\cal W}^{[0]}}{\delta p_{\alpha}(x)}
       = \frac{\delta \Gamma^{[0]}}{\delta p_{\alpha}(x)} \,, \ \ \ \ \ \
\\
&& \frac{\delta {\cal W}^{[0]}}{\delta n'_{\alpha}(x) }
       = \frac{\delta \Gamma^{[0]}}{\delta n'_{\alpha}(x) } \,, \ \ \ \ \ \
 \frac{\delta {\cal W}^{[0]}}{\delta l'_{\alpha}(x) }
       = \frac{\delta \Gamma^{[0]}}{\delta l'_{\alpha}(x) } \,, \ \ \ \ \ \
 \frac{\delta {\cal W}^{[0]}}{\delta p'_{\alpha}(x)}
       = \frac{\delta \Gamma^{[0]}}{\delta p'_{\alpha}(x)} \,. \ \ \ \ \ \
\end{eqnarray*}
\par
  In order to find the Slavnov--Taylor identity satisfied by the generating
functional for the regular vertex functions, we change the variables in the
path integral of ${\cal Z}^{[0]}$ as follows
\begin{eqnarray*}
&& W_a^{\mu}(x) \rightarrow
   W_a^{\mu}(x) + \delta \zeta \Delta W_a^{\mu}(x)\,, \ \ \ \ \
 W_y^{\mu}(x) \rightarrow
   W_y^{\mu}(x) + \delta \zeta \Delta W_y^{\mu}(x)\,, \\
&& C_a(x) \rightarrow
    C_a(x) + \delta \zeta \Delta C_a(x) \,, \ \ \ \ \
 C_y(x) \rightarrow C_y(x) \,,\\
&& \overline{C}_a(x) \rightarrow  \overline{C}_a(x)
                           - \delta \zeta \lambda_a(x)\,,\ \ \ \ \
 \overline{C}_y(x) \rightarrow  \overline{C}_y(x)
                           - \delta \zeta \lambda_y(x)\,, \\
&& \psi(x) \rightarrow \psi(x) +\delta \zeta \Delta \psi(x) \,, \ \ \ \ \
 \overline{\psi}(x) \rightarrow \overline{\psi}(x)
        + \delta \zeta \Delta \overline{\psi}(x) \,, \\
&& \lambda_a(x) \rightarrow  \lambda_a(x) \,, \ \ \ \ \
 \lambda_y(x) \rightarrow  \lambda_y(x) \,.
\end{eqnarray*}
The volume element of the path integral does not change and the changes in
$I_s$ and ${\cal L}_{WM}$ lead to
\begin{eqnarray}
&& \int d^4x \Big\{
\frac{\delta \Gamma^{[0]}} {\delta K^a_{\mu}(x)}
 \frac{\delta \Gamma^{[0]}} {\delta \widetilde{W}_a^{\mu}(x)} 
+\frac{\delta \Gamma^{[0]}} {\delta K^y_{\mu}(x)}
 \frac{\delta \Gamma^{[0]}} {\delta \widetilde{W}_y^{\mu}(x)}
+ \frac{\delta \Gamma^{[0]}} {\delta L_a(x)}
 \frac{\delta \Gamma^{[0]}} {\delta \widetilde{C}_a(x)} \nonumber\\
&&\ \ \ \ \ \ \ \ \ \  
+\frac{\delta \Gamma^{[0]}} {\delta \widetilde{\nu}_{L \alpha}(x)}
  \frac{\delta \Gamma^{[0]}} {\delta n_{\alpha}(x)} 
+ \frac{\delta \Gamma^{[0]}} {\delta \widetilde{e}_{L \alpha}(x)}
  \frac{\delta \Gamma^{[0]}} {\delta l_{\alpha}(x)} 
+ \frac{\delta \Gamma^{[0]}} {\delta \widetilde{e}_{R \alpha}(x)}
  \frac{\delta \Gamma^{[0]}} {\delta p_{\alpha}(x)} \nonumber\\
&&\ \ \ \ \ \ \ \ \ \ 
+\frac{\delta \Gamma^{[0]}} {\delta \widetilde{\overline{\nu}}_{L \alpha}(x)}
  \frac{\delta \Gamma^{[0]}} {\delta n'_{\alpha}(x)} 
+ \frac{\delta \Gamma^{[0]}} {\delta \widetilde{\overline{e}}_{L \alpha}(x)}
  \frac{\delta \Gamma^{[0]}} {\delta l'_{\alpha}(x)} 
+ \frac{\delta \Gamma^{[0]}} {\delta \widetilde{\overline{e}}_{R \alpha}(x)}
  \frac{\delta \Gamma^{[0]}} {\delta p'_{\alpha}(x)} \nonumber \\
&&\ \ \ \ \ \ \ \ \ \ 
 - \widetilde{\lambda}_a(x)
   \frac{\delta \Gamma^{[0]}} {\delta \widetilde{\overline{C}}_a(x)}
  - \widetilde{\lambda}_y(x) 
  + \frac{\delta \Gamma^{[0]}} {\delta \widetilde{\overline{C}}_y(x)}
  - \langle \Delta {\cal L}_{WM}(x) \rangle^{[0]}
 \Big\} = 0 \,,
\end{eqnarray}
where
$$
\langle \Delta {\cal L}_{WM}(x) \rangle^{[0]}
= \frac{1}{N{\cal Z}^{[0]}} \int {\cal D}
  [\overline{\psi},\psi,{\cal W}, \overline{C},C]
  \Delta {\cal L}_{WM}(x)
 {\rm exp} \Big\{ {\rm i} \big( I^{[0]}_{{\rm eff}} + I_s \big) \Big\}\,.
$$
With the definition of $\Delta {\cal L}_{WM}(x)$ 
$$
 \delta_B {\cal L}_{WM}(x) = \delta \zeta \Delta {\cal L}_{WM}(x) \,,
$$
one can write
\begin{eqnarray*}
 \langle \Delta {\cal L}_{WM}(x) \rangle^{[0]}
&=& M^2 \widetilde{W}_{a \mu}(x)
       \frac{\delta\Gamma^{[0]}}{\delta K^a_{\mu}(x)}
  + M^2 \Big(\frac{g_1}{g}\Big)^2 \widetilde{W}_{y \mu}(x)
       \frac{\delta\Gamma^{[0]}}{\delta K^y_{\mu}(x)} \\
  &&- M^2 \frac{g_1}{g} \widetilde{W}_{y \mu}(x)
       \frac{\delta\Gamma^{[0]}}{\delta K^3_{\mu}(x)}
   - M^2 \frac{g_1}{g} \widetilde{W}_{3 \mu}(x)
       \frac{\delta\Gamma^{[0]}}{\delta K^y_{\mu}(x)}  \,.
\end{eqnarray*}
Next, from the invariance of the path integral of ${\cal Z}^{[0]}$ with
respect to the translation of the integration variables $\overline{C}_a(x)$,
$\overline{C}_y(x)$, $\lambda_a(x)$ and $\lambda_y(x)$, one can get a set of
auxiliary identities
\begin{eqnarray}
&&   \frac{\delta \Gamma^{[0]}}
         {\delta \widetilde{\overline{C}}_1(x)}
   - \partial_{\mu}\frac{\delta \Gamma^{[0]}} {\delta K^1_{\mu}(x)}
   - g_1 \widetilde{W}_{y \mu}\frac{\delta \Gamma^{[0]}}
         {\delta K^2_{\mu}(x)}
   - g_1 \widetilde{W}_{2 \mu}\frac{\delta \Gamma^{[0]}}
         {\delta K^y_{\mu}(x)} = 0  \,, \\
&&   \frac{\delta \Gamma^{[0]}}
         {\delta \widetilde{\overline{C}}_2(x)}
   - \partial_{\mu}\frac{\delta \Gamma^{[0]}} {\delta K^2_{\mu}(x)}
   + g_1 \widetilde{W}_{y \mu} \frac{\delta \Gamma^{[0]}}
         {\delta K^1_{\mu}(x)}
   + g_1 \widetilde{W}_{1 \mu} \frac{\delta \Gamma^{[0]}}
         {\delta K^y_{\mu}(x)} = 0  \,,\\
&&   \frac{\delta \Gamma^{[0]}}
         {\delta \widetilde{\overline{C}}_3(x)}
   - \partial_{\mu}\frac{\delta \Gamma^{[0]}} {\delta K^3_{\mu}(x)} 
   - g_1 \widetilde{W}_{y \mu}\frac{\delta \Gamma^{[0]}}
         {\delta K^3_{\mu}(x)}
   - g_1 \widetilde{W}_{3 \mu}\frac{\delta \Gamma^{[0]}}
         {\delta K^y_{\mu}(x)} = 0  \,,\\
&&   \frac{\delta \Gamma^{[0]}}
         {\delta \widetilde{\overline{C}}_y(x)}
   - \partial_{\mu}\frac{\delta \Gamma^{[0]}} {\delta K^y_{\mu}(x)} 
   - g \widetilde{W}_{y \mu}\frac{\delta \Gamma^{[0]}}
         {\delta K^3_{\mu}(x)}
   - g \widetilde{W}_{3 \mu}\frac{\delta \Gamma^{[0]}}
         {\delta K^y_{\mu}(x)} = 0  \,,
\end{eqnarray}      
and
\begin{eqnarray}
 \frac{\delta \Gamma^{[0]}}{\delta \widetilde{\lambda}_a(x)}
                   = \langle \Phi_a(x) \rangle^{[0]}  \,,\ \ \ \ \
  \frac{\delta \Gamma^{[0]}}{\delta \widetilde{\lambda}_y(x)}
                   = \langle \Phi_y(x) \rangle^{[0]}  \,.
\end{eqnarray}      
where
\begin{eqnarray}
&& \langle \Phi_a(x) \rangle^{[0]}
  = \frac{1}{N{\cal Z}^{[0]}} \int {\cal D}
   [\overline{\psi},\psi,{\cal W}, \overline{C},C,\lambda] \Phi_a(x)
   {\rm exp} \Big\{ {\rm i} \big( I^{[0]}_{{\rm eff}} + I_s \big) \Big\}\,,
\\
&& \langle \Phi_y(x) \rangle^{[0]}
  = \frac{1}{N{\cal Z}^{[0]}} \int {\cal D}
   [\overline{\psi},\psi,{\cal W}, \overline{C},C,\lambda] \Phi_y(x)
   {\rm exp} \Big\{ {\rm i} \big( I^{[0]}_{{\rm eff}} + I_s \big) \Big\}\,.
\end{eqnarray}
Let $\widetilde{\Phi}_a(x) $, $\widetilde{\Phi}_y(x)$,
 $\widetilde{{\cal L}}_{WM}$ be the results obtained from $\Phi_a(x)$,
$\Phi_y(x)$, ${\cal L}_{WM}$ by replacing the field functions with the
classical field functions and define
\begin{eqnarray}
 \overline{\Gamma}^{[0]}
 = \Gamma^{[0]} - \int d^4x \Big\{
       \widetilde{\lambda}_a(x)\widetilde{\Phi}_a(x) 
      + \widetilde{\lambda}_y(x)\widetilde{\Phi}_y(x) 
      + \widetilde{{\cal L}}_{WM}
   \Big\} \,,
\end{eqnarray}
Thus, from (4.6)--(4.11), one gets
\begin{eqnarray}
&& \int d^4x \Big\{ 
\frac{\delta \overline{\Gamma}^{[0]}} {\delta K^a_{\mu}(x)}
 \frac{\delta \overline{\Gamma}^{[0]}} {\delta \widetilde{W}_a^{\mu}(x)} 
+\frac{\delta \overline{\Gamma}^{[0]}} {\delta K^y_{\mu}(x)}
 \frac{\delta \overline{\Gamma}^{[0]}} {\delta \widetilde{W}_y^{\mu}(x)}
+ \frac{\delta \overline{\Gamma}^{[0]}} {\delta L_a(x)}
 \frac{\delta \overline{\Gamma}^{[0]}} {\delta \widetilde{C}_a(x)}\nonumber\\
&&\ \ \ \ \ \ \ \ \ \ 
+\frac{\delta \overline{\Gamma}^{[0]}} {\delta \widetilde{\nu}_{L \alpha}(x)}
  \frac{\delta \overline{\Gamma}^{[0]}} {\delta n_{\alpha}(x)} 
+ \frac{\delta \overline{\Gamma}^{[0]}} {\delta \widetilde{e}_{L \alpha}(x)}
  \frac{\delta \overline{\Gamma}^{[0]}} {\delta l_{\alpha}(x)} 
+ \frac{\delta \overline{\Gamma}^{[0]}} {\delta \widetilde{e}_{R \alpha}(x)}
  \frac{\delta \overline{\Gamma}^{[0]}} {\delta p_{\alpha}(x)} \nonumber\\
&&\ \ \ \ \ \ \ \ \ \ 
+\frac{\delta \overline{\Gamma}^{[0]}}
       {\delta \widetilde{\overline{\nu}}_{L \alpha}(x)}
  \frac{\delta \overline{\Gamma}^{[0]}} {\delta n'_{\alpha}(x)} 
+ \frac{\delta \overline{\Gamma}^{[0]}}
       {\delta \widetilde{\overline{e}}_{L \alpha}(x)}
  \frac{\delta \overline{\Gamma}^{[0]}} {\delta l'_{\alpha}(x)} 
+ \frac{\delta \overline{\Gamma}^{[0]}}
       {\delta \widetilde{\overline{e}}_{R \alpha}(x)}
  \frac{\delta \overline{\Gamma}^{[0]}} {\delta p'_{\alpha}(x)} 
 \Big\} = 0 \,.
\end{eqnarray}
and
\begin{eqnarray}
&& \frac{\delta \overline{\Gamma}^{[0]}}{\delta \widetilde{\lambda}_a(x)}
    = \langle \Phi_a(x) \rangle^{[0]} - \widetilde{\Phi}_a(x)  \,,\ \ \ \ \
  \frac{\delta \overline{\Gamma}^{[0]}}{\delta \widetilde{\lambda}_y(x)}
    = \langle \Phi_y(x) \rangle^{[0]} - \widetilde{\Phi}_y(x) \,,\\
&& \frac{\delta \overline{\Gamma}^{[0]}}
         {\delta \widetilde{\overline{C}}_1(x)}
 - \partial_{\mu}\frac{\delta \overline{\Gamma}^{[0]}} {\delta K^1_{\mu}(x)}
 - g_1 \widetilde{W}_{y \mu}\frac{\delta \overline{\Gamma}^{[0]}}
         {\delta K^2_{\mu}(x)}
 - g_1 \widetilde{W}_{2 \mu}\frac{\delta \overline{\Gamma}^{[0]}}
         {\delta K^y_{\mu}(x)} = 0  \,,\\
&& \frac{\delta \overline{\Gamma}^{[0]}}
         {\delta \widetilde{\overline{C}}_2(x)}
 - \partial_{\mu}\frac{\delta \overline{\Gamma}^{[0]}} {\delta K^2_{\mu}(x)}
 + g_1 \widetilde{W}_{y \mu} \frac{\delta \overline{\Gamma}^{[0]}}
         {\delta K^1_{\mu}(x)}
 + g_1 \widetilde{W}_{1 \mu} \frac{\delta \overline{\Gamma}^{[0]}}
         {\delta K^y_{\mu}(x)} = 0  \,,\\
&& \frac{\delta \overline{\Gamma}^{[0]}}
         {\delta \widetilde{\overline{C}}_3(x)}
 - \partial_{\mu}\frac{\delta \overline{\Gamma}^{[0]}} {\delta K^3_{\mu}(x)} 
 - g_1 \widetilde{W}_{y \mu}\frac{\delta \overline{\Gamma}^{[0]}}
         {\delta K^3_{\mu}(x)}
 - g_1 \widetilde{W}_{3 \mu}\frac{\delta \overline{\Gamma}^{[0]}}
         {\delta K^y_{\mu}(x)} = 0  \,,\\
&& \frac{\delta \overline{\Gamma}^{[0]}}
         {\delta \widetilde{\overline{C}}_y(x)}
 - \partial_{\mu}\frac{\delta \overline{\Gamma}^{[0]}} {\delta K^y_{\mu}(x)} 
 - g \widetilde{W}_{y \mu}\frac{\delta \overline{\Gamma}^{[0]}}
         {\delta K^3_{\mu}(x)}
 - g \widetilde{W}_{3 \mu}\frac{\delta \overline{\Gamma}^{[0]}}
         {\delta K^y_{\mu}(x)} = 0  \,.
\end{eqnarray}      
    As $\Phi_a(x), \Phi_y(x)$ contain the products of the SU$_L$(2) and
U$_Y$(1) fields, (4.16) is complicated unless the generating functional for
the Green functions does not contain the sources of the Lagrange multipliers
$\lambda_a$ and $\lambda_y$. Actuarely we are not interested in using the
Green functions involving $\lambda_a$ or $\lambda_y$. Our intention to use
the generalized form of the theory containing the sources of these Lagrange
multipliers is to study the Renormalisability of the theory for which such
sources are absent from the generating functional for the Green functions
and therefore $\langle \Phi_a(x) \rangle^{[0]}$ and
$\langle \Phi_y(x) \rangle^{[0]}$ are equal to zero. We now, according to
$(4.11)$, let vanish
$\frac{\delta \Gamma^{[0]}}{\delta \widetilde{\lambda}_a(x)}$ and
$\frac{\delta \Gamma^{[0]}}{\delta \widetilde{\lambda}_a(x)}$ to make
$\langle \Phi_a(x) \rangle^{[0]}$ and $\langle \Phi_y(x) \rangle^{[0]}$
equal to zero. This means
\begin{eqnarray}
 \widetilde{\Phi}_a(x) = 0 \,,\ \ \ \ \
  \widetilde{\Phi}_y(x) = 0 \,.
\end{eqnarray} 
and
\begin{eqnarray}
 \frac{\delta \overline{\Gamma}^{[0]}}{\delta \widetilde{\lambda}_a(x)}
    = 0 \,,\ \ \ \ \
  \frac{\delta \overline{\Gamma}^{[0]}}{\delta \widetilde{\lambda}_y(x)}
    = 0 \,,
\end{eqnarray} 
\par
      In the following we will denote by $\overline{\Gamma}^{[0]}
 [\psi,\overline{\psi},W,\overline{C}, C, \lambda,K,L,n,l,p,n',l',p']$
the functional that is obtained from $\overline{\Gamma}^{[0]}
 [\widetilde{\psi},\widetilde{\overline{\psi}},\widetilde{W},
  \widetilde{\overline{C}}, \widetilde{C}, \widetilde{\lambda},K,\cdots]$
by replacing the classical field functions with the usual field functions.
Assume that the dimensional regularization method is used and the
Slavnov--Taylor identity and the auxiliary identities are guaranteed.
 Denote the tree part and one loop part of $\overline{\Gamma}^{[0]}$ by
$\overline{\Gamma}^{[0]}_0$ and $\overline{\Gamma}^{[0]}_1$  respectively.
 $\overline{\Gamma}^{[0]}_0$ is thus the modified action
$\overline{I}^{[0]}_{eff}$ obtained from $I^{[0]}_{eff}$ by excluding the
mass term and $(\lambda_a,\lambda_y)$ terms. From $(4.15)$
and $(4.17)-(4.22)$ one has
\begin{eqnarray}
&& \Phi_a(x) = 0 \,,\ \ \ \ \  \Phi_y(x) = 0 \,,\\
&& \frac{\delta \overline{\Gamma}^{[0]}}{\delta \lambda_a(x)}
    = 0 \,,\ \ \ \ \
  \frac{\delta \overline{\Gamma}^{[0]}}{\delta \lambda_y(x)} = 0 \,,\\
&& \Lambda_{op} \overline{\Gamma}^{[0]}_0 = 0 \,,  \nonumber
\end{eqnarray}
and
\begin{eqnarray}
&& \overline{\Gamma}^{[0]}_0 * \overline{\Gamma}^{[0]}_1
  + \overline{\Gamma}^{[0]}_1 * \overline{\Gamma}^{[0]}_0
  = \Lambda_{op} \overline{\Gamma}^{[0]}_1 = 0 \,, \\
&& \Sigma_a(x) \overline{\Gamma}^{[0]} = 0 \,, \ \ \ \ \ \
  \Sigma_y(x) \overline{\Gamma}^{[0]} = 0 \,.
\end{eqnarray}
where $\Lambda_{op}$,$\Sigma_a(x)$ and $\Sigma_y(x)$ are defined by
\begin{eqnarray}
&& \Lambda_{op} =  \int d^4x \Big\{ 
   \frac{\delta \overline{\Gamma}^{[0]}_0} {\delta K^a_{\mu}(x)}
   \frac{\delta } {\delta W_a^{\mu}(x)} 
 + \frac{\delta \overline{\Gamma}^{[0]}_0} {\delta W_a^{\mu}(x)}
   \frac{\delta } {\delta K^a_{\mu}(x)} 
 + \frac{\delta \overline{\Gamma}^{[0]}_0} {\delta K^y_{\mu}(x)}
   \frac{\delta } {\delta W_y^{\mu}(x)}
 + \frac{\delta \overline{\Gamma}^{[0]}_0} {\delta W_y^{\mu}(x)}
   \frac{\delta } {\delta K^y_{\mu}(x)} \nonumber\\
&&\ \ \ \ \ \ \ + \frac{\delta \overline{\Gamma}^{[0]}_0} {\delta L_a(x)}
   \frac{\delta } {\delta C_a(x)}
 + \frac{\delta \overline{\Gamma}^{[0]}_0} {\delta C_a(x)} 
   \frac{\delta } {\delta L_a(x)} 
 + \frac{\delta \overline{\Gamma}^{[0]}_0} {\delta \nu_{L \alpha}(x)}
   \frac{\delta } {\delta n_{\alpha}(x)}
 + \frac{\delta \overline{\Gamma}^{[0]}_0} {\delta n_{\alpha}(x)}
   \frac{\delta } {\delta \nu_{L \alpha}(x)}\nonumber\\
&&\ \ \ \ \ \ \ + \frac{\delta \overline{\Gamma}^{[0]}_0}
   {\delta e_{L \alpha}(x)}\frac{\delta } {\delta l_{\alpha}(x)} 
 + \frac{\delta \overline{\Gamma}^{[0]}_0} {\delta l_{\alpha}(x)}
   \frac{\delta } {\delta e_{L \alpha}(x)} 
 + \frac{\delta \overline{\Gamma}^{[0]}_0} {\delta e_{R \alpha}(x)}
   \frac{\delta } {\delta p_{\alpha}(x)} 
 + \frac{\delta \overline{\Gamma}^{[0]}_0} {\delta p_{\alpha}(x)}
   \frac{\delta } {\delta e_{R \alpha}(x)} \nonumber\\
&&\ \ \ \ \ \ \ + \frac{\delta \overline{\Gamma}^{[0]}_0}
       {\delta \overline{\nu}_{L \alpha}(x)}
   \frac{\delta } {\delta n'_{\alpha}(x)} 
 + \frac{\delta \overline{\Gamma}^{[0]}_0} {\delta n'_{\alpha}(x)} 
   \frac{\delta } {\delta \overline{\nu}_{L \alpha}(x)} 
 + \frac{\delta \overline{\Gamma}^{[0]}_0}
       {\delta \overline{e}_{L \alpha}(x)}
   \frac{\delta } {\delta l'_{\alpha}(x)} 
 + \frac{\delta \overline{\Gamma}^{[0]}_0} {\delta l'_{\alpha}(x)} 
   \frac{\delta } {\delta \overline{e}_{L \alpha}(x)} \nonumber\\
&&\ \ \ \ \ \ \  + \frac{\delta \overline{\Gamma}^{[0]}_0}
       {\delta \overline{e}_{R \alpha}(x)}
   \frac{\delta } {\delta p'_{\alpha}(x)} 
 + \frac{\delta \overline{\Gamma}^{[0]}_0} {\delta p'_{\alpha}(x)}
   \frac{\delta } {\delta \overline{e}_{R \alpha}(x)} \Big\} \,, \\
&& \Sigma_1(x) =  \frac{\delta}{\delta \overline{C}_1(x)}
       - \partial_{\mu}\frac{\delta}{\delta K^1_{\mu}(x)}
       - g_1 W_{y \mu}\frac{\delta}{\delta K^2_{\mu}(x)}
       - g_1 W_{2 \mu}\frac{\delta}{\delta K^y_{\mu}(x)}   \,,\\
&& \Sigma_2(x) =  \frac{\delta}{\delta \overline{C}_2(x)}
       - \partial_{\mu}\frac{\delta}{\delta K^2_{\mu}(x)}
       + g_1 W_{y \mu}\frac{\delta}{\delta K^1_{\mu}(x)}
       + g_1 W_{1 \mu}\frac{\delta}{\delta K^y_{\mu}(x)}   \,,\\
&& \Sigma_3(x) =  \frac{\delta}{\delta \overline{C}_3(x)}
       - \partial_{\mu}\frac{\delta}{\delta K^3_{\mu}(x)}
       - g_1 W_{y \mu}\frac{\delta}{\delta K^3_{\mu}(x)}
       - g_1 W_{3 \mu}\frac{\delta}{\delta K^y_{\mu}(x)}   \,,  \\
&& \Sigma_y(x) =  \frac{\delta}{\delta \overline{C}_y(x)}
       - \partial_{\mu}\frac{\delta}{\delta K^y_{\mu}(x)}
       - g W_{y \mu}\frac{\delta}{\delta K^3_{\mu}(x)}
       - g W_{3 \mu}\frac{\delta}{\delta K^y_{\mu}(x)}   \,.
\end{eqnarray}
The meaning of the notation $A*B$ is the same as in the common use, namely
\begin{eqnarray}
&&A*B = \int d^4x \Big\{
\frac{\delta A} {\delta K^a_{\mu}(x)}
 \frac{\delta B} {\delta W_a^{\mu}(x)} 
+\frac{\delta A} {\delta K^y_{\mu}(x)}
 \frac{\delta B} {\delta W_y^{\mu}(x)}
+ \frac{\delta A} {\delta L_a(x)}
 \frac{\delta B} {\delta C_a(x)} \nonumber\\
&&\ \ \ \ \ \ \ \ \ \ 
+\frac{\delta A} {\delta \nu_{L \alpha}(x)}
  \frac{\delta B} {\delta n_{\alpha}(x)} 
+ \frac{\delta A} {\delta e_{L \alpha}(x)}
  \frac{\delta B} {\delta l_{\alpha}(x)} 
+ \frac{\delta A} {\delta e_{R \alpha}(x)}
  \frac{\delta B} {\delta p_{\alpha}(x)} \nonumber\\
&&\ \ \ \ \ \ \ \ \ \ 
+\frac{\delta A} {\delta \overline{\nu}_{L \alpha}(x)}
  \frac{\delta B} {\delta n'_{\alpha}(x)} 
+ \frac{\delta A} {\delta \overline{e}_{L \alpha}(x)}
  \frac{\delta B} {\delta l'_{\alpha}(x)} 
+ \frac{\delta A} {\delta \overline{e}_{R \alpha}(x)}
  \frac{\delta B} {\delta p'_{\alpha}(x)} 
 \Big\}  \,.
\end{eqnarray}  
$(4.24)-(4.26)$ are of course satisfied by the finite part and the pole part
of $\overline{\Gamma}^{[0]}_1$. Thus the equations of the pole part
 $\overline{\Gamma}^{[0]}_{1,div}$ are
\begin{eqnarray}
&& \frac{\delta \overline{\Gamma}^{[0]}_{1,div}}{\delta \lambda_a(x)}
    = 0 \,,\ \ \ \ \
  \frac{\delta \overline{\Gamma}^{[0]}_{1,div}}{\delta \lambda_y(x)}
    = 0 \,, \\
&&\Lambda_{op} \overline{\Gamma}^{[0]}_{1,div} = 0 \,,\\
&& \Sigma_a(x) \overline{\Gamma}^{[0]}_{1,div} = 0 \,, \ \ \ \ \ \
  \Sigma_y(x) \overline{\Gamma}^{[0]}_{1,div} = 0 \,.
\end{eqnarray}
Obviously, the same equations should be found for a SU$_L$(2)$\times$U$_Y$(1)
theory without the mass term if the same constraint conditions are chosen. 
\par
    If $M=0$, then it is known from the renormalisability of the theory that
$\overline{\Gamma}^{[0]}_{1,div} $ is a combination of the following terms
\begin{eqnarray*}
&& T_{GL} = g \frac{\partial \overline{\Gamma}^{[0]}_0}{\partial g} \,,\ \ \ \
T_{GY} = g_1 \frac{\partial \overline{\Gamma}^{[0]}_0}{\partial g_1} \,,\\
&& T_{WL} = \int d^4x \Big\{
   W_a^{\mu}(x) \frac{\delta \overline{\Gamma}^{[0]}_0}
                                 {\delta W_a^{\mu}(x)} 
  + L_a(x) \frac{\delta \overline{\Gamma}^{[0]}_0} {\delta L_a(x)}
 \Big\} \,,\\
&& T_{WY} = \int d^4x 
   W_y^{\mu}(x) \frac{\delta \overline{\Gamma}^{[0]}_0}
                                 {\delta W_y^{\mu}(x)} \,,\\
&& T_{CK} = \int d^4x \Big\{
   \overline{C}_a(x) \frac{\delta \overline{\Gamma}^{[0]}_0}
                                 {\delta \overline{C}_a(x)} 
 + C_a(x) \frac{\delta \overline{\Gamma}^{[0]}_0}
                           {\delta C_a(x)} 
 + K^a_{\mu}(x) \frac{\delta \overline{\Gamma}^{[0]}_0}
                     {\delta K^a_{\mu}(x)}
 \Big\} \,, \\
&& T_{CKY} = \int d^4x \Big\{
   \overline{C}_y(x) \frac{\delta \overline{\Gamma}^{[0]}_0}
                                 {\delta \overline{C}_(x)} 
 + C_y(x) \frac{\delta \overline{\Gamma}^{[0]}_0}
                           {\delta C_y(x)} 
 + K^y_{\mu}(x) \frac{\delta \overline{\Gamma}^{[0]}_0}
                     {\delta K^y_{\mu}(x)}
 \Big\} \,, \\
&& T_{\nu L}= \int d^4x \Big\{
     \nu_{L \alpha}(x) \frac{\delta \overline{\Gamma}^{[0]}_0}
                                   {\delta \nu_{L \alpha}(x)}
     + \overline{\nu}_{L \alpha}(x)
            \frac{\delta \overline{\Gamma}^{[0]}_0}
                 {\delta \overline{\nu}_{L \alpha}(x)} \Big\} \,,\\
&& T_{eL} = \int d^4x \Big\{
     e_{L \alpha}(x) \frac{\delta \overline{\Gamma}^{[0]}_0}
                                   {\delta e_{L \alpha}(x)}
    + \overline{e}_{L \alpha}(x)
            \frac{\delta \overline{\Gamma}^{[0]}_0}
                 {\delta \overline{e}_{L \alpha}(x)} \Big\} \,, \\
&& T_{eR} = \int d^4x \Big\{
      e_{R \alpha}(x) \frac{\delta \overline{\Gamma}^{[0]}_0}
                                   {\delta e_{R \alpha}(x)}
     + \overline{e}_{R \alpha}(x)
            \frac{\delta \overline{\Gamma}^{[0]}_0}
                 {\delta \overline{e}_{R \alpha}(x)} \Big\} \,,\\
&& T_{nn'} = \int d^4x \Big\{
          n_{\alpha}(x) \frac{\delta \overline{\Gamma}^{[0]}_0}
                                   {\delta n_{\alpha}(x)}
          + n'_{\alpha}(x) \frac{\delta \overline{\Gamma}^{[0]}_0}
                                   {\delta n'_{\alpha}(x)}  \Big\} \,, \\
&& T_{ll'} = \int d^4x \Big\{
           l_{\alpha}(x) \frac{\delta \overline{\Gamma}^{[0]}_0}
                                     {\delta l_{\alpha}(x)}
           + l'_{\alpha}(x) \frac{\delta \overline{\Gamma}^{[0]}_0}
                                     {\delta l'_{\alpha}(x)} \Big\} \,,\\
&& T_{pp'} = \int d^4x \Big\{
            p_{\alpha}(x) \frac{\delta \overline{\Gamma}^{[0]}_0}
                                      {\delta p_{\alpha}(x)}
            + p'_{\alpha}(x) \frac{\delta \overline{\Gamma}^{[0]}_0}
                                   {\delta p'_{\alpha}(x)} \Big\}  \,.
\end{eqnarray*}
With these terms one can form five solutions of equations $(4.33)-(4.35)$,
which can be chosen as
\begin{eqnarray}
&& T_{(1)} = T_{WL} - T_{GL} - T_{CK} \,,\\
&& T_{(2)} = T_{WY} - T_{GY} - T_{CKY} \,,\\
&& T_{(3)} = T_{CK} + T_{CKY} + T_{nn'} + T_{ll'} + T_{pp'}  \,,\\
&& T_{(4)} = T_{\nu L} + T_{eL} - T_{nn'} - T_{ll'} \,, \\
&& T_{(5)} = T_{eR} - T_{pp'}  \,.
\end{eqnarray}
\par
Note that $T_{(3)}$ is
$2\big(\overline{\Gamma}^{[0]}_0 -I_{WL}-I_{WY}-I_{\psi}-I_{\psi W}\big)$.
$T_{(1)}$ is a combination of $I_{WL}$, $T_{(3)}$ and
$\int d^4x C_y(x) \frac{\delta \overline{\Gamma}^{[0]}_0}{\delta C_y(x)}$. 
$T_{(2)}$ is a combination of $I_{WY}$ and
$\int d^4x C_y(x) \frac{\delta \overline{\Gamma}^{[0]}_0}{\delta C_y(x)}$.
The sum of $T_{(4)}$ and $T_{(5)}$ is $2\big(I_{\psi}+I_{\psi W}\big)$. 
 $\int d^4x C_y(x) \frac{\delta \overline{\Gamma}^{[0]}_0}{\delta C_y(x)}$
and $T_{(5)}$ can be easily checked to satisfy $(4.34)-(4.35)$. In addition
to $(4.36)-$$(4.40)$, a new term appearing
in $\overline{\Gamma}^{[0]}_{1,\rm div}$ when $M\not= 0$ should includ $M^2$
as a factor and also satisfies $(4.34)-(4.35)$. Only $I_{WM}$ can be a
candidate. Is such a term can really appear ? Imagine a limiting case that
the matter fields and the U$_Y(1)$ fields are absent. Thus the constraint
conditions become Lorentz conditions and the above five solutions become two,
namely, $\big(T_{WL} - T_{GL} \big)$ and $T_{CK}$. This combination of
$T_{WL}$ and $T_{GL}$ are due to the restriction of the constraint condition
containing $\partial^{\mu}W_{y\mu}$ and therefore should be decomposed into
two independent terms when the U$_Y(1)$ fields are absent. In fact, it is
known that a SU(n) theory with massive gauge Bosons is
renormalisability [1] and that when the matter fields are absent
 $\overline{\Gamma}^{[n]}_{n+1,div}$ of such a theory is a combination
of three independent terms $T_{WL}$, $T_{GL}$ and $T_{CK}$. It follows that
for the present theory $\overline{\Gamma}^{[0]}_{1,div}$ does not cantain
the mass term $I_{MW}$ neither and can be expressed as
\begin{eqnarray}
\overline{\Gamma}^{[0]}_{1,div}
 =\, \alpha_1^{(1)}T_{(1)} + \alpha_2^{(1)}T_{(2)} + \alpha_3^{(1)}T_{(3)}
    + \alpha_4^{(1)}T_{(4)} + \alpha_5^{(1)} T_{(5)} \,,
\end{eqnarray}
where, $\alpha_1^{(1)}, \cdots, \alpha_5^{(1)}$ are constants of order
 $(\hbar)^1$ and are divergent when the space-time dimension tends to $4$.
\par
    In order to cancel the one loop divergence the counterterm of order
$\hbar^1$ in the action should be chosen as 
\begin{eqnarray}
 \delta I^{[1]}_{count} = - \overline{\Gamma}^{[0]}_{1,div} \,,
\end{eqnarray}
Since 
\begin{eqnarray}
 \overline{I}_{eff}^{[0]} = \overline{\Gamma}^{[0]}_0 \,,
\end{eqnarray}
it is known from $(4.41)$ that the sum of $\overline{I}_{eff}^{[0]}$ and
$\delta I^{[1]}_{count}$, to order of $\hbar^1$, can be written as
\begin{eqnarray}
 \overline{I}_{eff}^{[1]}
  && [\psi,\overline{\psi},W,C,\overline{C},
  K,L,n,l,p,n',l',p', g,g_1] \nonumber\\
 && = \overline{I}_{{\rm eff}}^{[0]}[\psi^{[0]},\overline{\psi}^{[0]},
 W^{[0]},C^{[0]},\overline{C}^{[0]}, K^{[0]}, L^{[0]},n^{[0]},n^{'[0]},
   \cdots, g^{[0]}, g_1^{[0]}]  \,,
\end{eqnarray}
where the bare fields and the bare parameters (to order $(\hbar)^1$) are
defined as
\begin{eqnarray}
&& W^{[0]}_{a \mu} = (Z_3^{[1]})^{1/2} W_{a \mu}
        =\big(1 - \alpha_1^{(1)}\big) W_{a \mu}\,,\ \ \ \
  L^{[0]}_a = (Z_3^{[1]})^{1/2} L_a  \,, \\
&&  W^{[0]}_{y \mu} = (Z_3^{'[1]})^{1/2} W_{y \mu}
        = \big(1 - \alpha_2^{(1)}\big) W_{y \mu}\,,\\
&& C^{[0]}_a = (\widetilde{Z}_3^{[1]})^{1/2}C_a
       = \big( 1 - \alpha_3^{(1)} + \alpha_1^{(1)} \big) C_a \,,\\
&& \overline{C}^{[0]}_a = (\widetilde{Z}_3^{[1]})^{1/2} \overline{C}_a \,,
 \ \ \ \ \
 K^{a[0]}_{\mu} = (\widetilde{Z}_3^{[1]})^{1/2} K^a_{\mu} \,,\\
&& C^{[0]}_y = (\widetilde{Z}_3^{'[1]})^{1/2}C_y
       = \big(1 - \alpha_3^{(1)} + \alpha_2^{(1)} \big) C_y \,,\\ 
&& \overline{C}^{[0]}_y = (\widetilde{Z}_3^{'[1]})^{1/2} \overline{C}_y \,,
  \ \ \ \
  K^{y[0]}_{\mu} = (\widetilde{Z}_3^{[1]})^{1/2} K^y_{\mu}\,,\\
&& \nu_L^{[0]} = (Z_{\nu L}^{[1]})^{1/2} \nu_L
         = \big(1 - \alpha_4^{(1)}\big) \nu_L \,,\ \ \ \
  \overline{\nu}_L^{[0]}
                 = (Z_{\nu L}^{[1]})^{1/2} \overline{\nu}_L \,,\\
&& e_L^{[0]} = (Z_{eL}^{[1]})^{1/2} e_L
            = (Z_{\nu L}^{[1]})^{1/2}e_L \,,\ \ \ \
   \overline{e}_L^{[0]}
                        = (Z_{eL}^{[1]})^{1/2} \overline{e}_L \,,\\
&& e_R^{[0]} = (Z_{eR}^{[1]})^{1/2} e_R
        = \big(1 - \alpha_5^{(1)}\big)e_R \,,\ \ \ \
  \overline{e}_R^{[0]}
                     = (Z_{eR}^{[1]})^{1/2} \overline{e}_R  \,,\\
&&  n^{[0]} = (Z_{(n)}^{[1]})^{1/2} n
      = \big(1 - \alpha_3^{(1)} + \alpha_4^{(1)} \big) n \,,\ \ \ \
    n^{'[0]} = (Z_{(n)}^{[1]})^{1/2} n' \,,\\
&& l^{[0]}  = (Z_{(l)}^{[1]})^{1/2} l
           = (Z_{(n)}^{[1]})^{1/2} l\,, \ \ \ \
    l^{'[0]} = (Z_{(l)}^{[1]})^{1/2} l' \,,\\
&&  p^{[0]}  = (Z_{(p)}^{[1]})^{1/2} p
     = \big( 1 - \alpha_3^{(1)} + \alpha_5^{(1)} \big) p \,,\ \ \ \
   p^{'[0]} = (Z_{(p)}^{[1]})^{1/2} p'  \,,\\
&&  g^{[0]} = Z_g^{[1]} g = (Z_3^{[1]})^{-1/2} g \,,\ \ \ \
   g_1^{[0]} = Z_g^{'[1]} g_1 = (Z_3^{'[1]})^{-1/2} g_1 \,.
\end{eqnarray}
\par
Next, defined
\begin{eqnarray*}
&& \Phi_1^{[0]}= \partial^{\mu}W_{1 \mu}^{[0]}
               + g_1^{[0]} W_{2 \mu}^{[0]} W_y^{\mu [0]}  \,,\\
&& \Phi_2^{[0]}= \partial^{\mu}W_{2 \mu}^{[0]}
               - g_1^{[0]} W_{1 \mu}^{[0]} W_y^{\mu [0]}  \,,\\
&& \Phi_3^{[0]}= \partial^{\mu}W_{3 \mu}^{[0]}
               + g_1^{[0]} W_{3 \mu}^{[0]} W_y^{\mu [0]}  \,, \\
&& \Phi_y^{[0]}= \partial^{\mu}W_{y \mu}^{[0]}
               + g^{[0]} W_{3 \mu}^{[0]} W_y^{\mu [0]} \,.
\end{eqnarray*}
From $(4.45),(4.46)$ and $(4.57)$ one has
$$
 g^{[0]} W_a^{\mu [0]} = g W_a^{\mu} \,, \ \ \ \ \ \
  g_1^{[0]} W_y^{\mu [0]} = g_1 W_y^{\mu} \,,
$$
and
\begin{eqnarray}
 \Phi_a^{[0]} = (Z_3^{[1]})^{1/2} \Phi_a  \,, \ \ \ \
  \Phi_y^{[0]} = (Z_3^{'[1]})^{1/2} \Phi_y  \,.
\end{eqnarray}
Thus by adding $I_{WM}$ and the $\lambda$ terms into
$\overline{I}_{eff}^{[1]}$ and forming
\begin{eqnarray} 
 I_{eff}^{[1]} 
 =\, \overline{I}_{eff}^{[1]} + I_{WM}
    + \int d^4x \Big\{
        \lambda_a(x) \Phi_a(x) + \lambda_y(x) \Phi_y(x) \Big\}\,,
\end{eqnarray}
one gets
\begin{eqnarray}
&& I_{eff}^{[1]}
  [\psi,\overline{\psi},W,C,\overline{C}, \lambda,
     K,L,n,l,p,n',l',p', g,g_1, M] \nonumber\\
 && \ \ \ \ \ 
  = I_{eff}^{[0]}[\psi^{[0]},\overline{\psi}^{[0]},
      W^{[0]},C^{[0]},\overline{C}^{[0]},\lambda^{[0]},
   K^{[0]}, L^{[0]},n^{[0]},n^{'[0]},
   \cdots, g^{[0]}, g_1^{[0]}, M^{[0]}] \,,
\end{eqnarray}
where 
\begin{eqnarray}
 M^{[0]} = (Z_3^{[1]})^{-1/2} M \,, \ \ \ \
  \lambda_a^{[0]} = (Z_3^{[1]})^{-1/2} \lambda_a \,, \ \ \ \
  \lambda_a^{[0]} = (Z_3^{'[1]})^{-1/2} \lambda_y \,.
\end{eqnarray}
Obviously, if the action $I_{eff}^{[1]}$ is used to replace $I_{eff}^{[0]}$
in $(4.3)$ and define  ${\cal Z}^{[1]}$, $\Gamma^{[1]}$ as well as
\begin{eqnarray}
\overline{\Gamma}^{[1]}
 = \Gamma^{[1]} - I_{WM} - \int d^4x \Big\{ \lambda_a(x)\Phi_a(x) 
                     + \lambda_y(x)\Phi_y(x) + {\cal L}_{WM} \Big\} \,,
\end{eqnarray}
then one has
\begin{eqnarray}
&&  \overline{\Gamma}^{[1]}[\psi,\overline{\psi},W,C,\overline{C},
  \lambda,K,L,n,l,p,n',l',p', g,g_1, M] \nonumber\\
&& \ \ \ \ \ \ 
= \overline{\Gamma}^{[0]}[\psi^{[0]},\overline{\psi}^{[0]},W^{[0]},
    C^{[0]},\overline{C}^{[0]},\lambda^{[0]},K^{[0]}, L^{[0]},
    n^{[0]},n^{'[0]},\cdots, g^{[0]}, g_1^{[0]}, M^{[0]}] \,.
\end{eqnarray}
From this it is easy to check that, to order $\hbar^1$,
$\overline{\Gamma}^{[1]}$ is finite. Moreover, by changing into bare fields
and bare parameters the fields and parameters in  $(4.15)-$ $(4.22)$  and
then transforming them back into the renormalized fields and renormalized
parameters according to $(4.45)-$$(4.59)$, one can see that, under
condition $(4.23)$, $\overline{\Gamma}^{[1]}$ also satisfies
\begin{eqnarray}
&& \Lambda_{op} \overline{\Gamma}^{[1]} = 0 \,,\\
&& \frac{\delta \overline{\Gamma}^{[1]}}{\delta \lambda_a(x)}
    = 0 \,,\ \ \ \ \
  \frac{\delta \overline{\Gamma}^{[1]}}{\delta \lambda_y(x)} = 0 \,,\\
&& \Sigma_a(x) \overline{\Gamma}^{[1]} = 0 \,, \ \ \ \ \ \
  \Sigma_y(x) \overline{\Gamma}^{[1]} = 0 \,.
\end{eqnarray}
\par
       It is now clear that the renormalisability  of the theory can be
verified by the inductive method. The following is an outline of the proof.
Assume that up to $n$ loop the theory has been proved to be renormalisable
by introducing the counterterm
$$
 I^{[n]}_{\rm count} = \sum_{l=1}^{n} \delta I^{[l]}_{\rm count}, 
$$
where $\delta I^{[l]}_{count}$ is the counterterm of order $\hbar^l$
and has the form of (4.41),(4.42).
Therefore the modified generating functional $\overline{\Gamma}^{[n]}$ for
the regular vertex, defined by the action 
$$ 
 I^{[n]}_{\rm eff} = I^{[0]}_{\rm eff} + I^{[n]}_{\rm count} 
$$
satisfied equations $(4.64)-(4.66)$ (under $(4.23)$) and, to order $\hbar^n$,
is finite. This also means that the fields or parameters in each of the
following brackets have the same renormalization factor:
$$
 ( W^{[0]}_{a \mu}, L_a), (C_a,\overline{C}_a,K^a_{\mu}),
(C_y,\overline{C}_y,K^y_{\mu}), ({\nu}_L,\overline{\nu}_L,
e_L, \overline{e}_L), (e_R,\overline{e}_R ), (n,n',l,l'), (p,p'),
(\lambda,M,g),
$$
and that
\begin{eqnarray*}
&& Z_g^{'[n]} (Z_3^{'[n]})^{1/2} = 1 \,, \ \ \ \ \
  Z_g^{[n]} (Z_3^{[n]})^{1/2} = 1 \,, \\
&& Z_3^{[n]} \widetilde{Z}_3^{[n]}
= \widetilde{Z}_3^{'[n]} \widetilde{Z}_3^{'[n]}
=  Z_{\nu L}^{[n]} Z_{(n)}^{[n]}
=  Z_{eR}^{[n]} Z_{(p)}^{[n]} \,.
\end{eqnarray*}
We have to proved that by using a counterterm of order $\hbar^{n+1}$
which also has the form of (4.41),(4.42), $\overline{\Gamma}^{[n+1]}$
can be make satisfy $(4.64)-(4.66)$ and finite to order $\hbar^{n+1}$,
where $\overline{\Gamma}^{[n+1]}$ is the modified generating functional for
the regular vertex, determined by the action 
$$ 
 I^{[n+1]}_{\rm eff} = I^{[n]}_{\rm eff} + \delta I^{[n+1]}_{\rm count}. 
$$                           
\par
    Denote by $\overline{\Gamma}^{[n]}_k$ the part of order $\hbar^{k}$
in $\overline{\Gamma}^{[n]}$. For $k\leq n$, $\overline{\Gamma}^{[n]}_k$ is
equal to $\overline{\Gamma}^{[k]}_k$, because it can not contain the
contribution of a counterterm of order $\hbar^{k+1}$ or higher. Thus
on expanding $\overline{\Gamma}^{[n]}$ to order $\hbar^{n+1}$ one has
$$
 \overline{\Gamma}^{[n]}
 = \sum_{k=0}^{n} \overline{\Gamma}^{[k]}_k + \overline{\Gamma}^{[n]}_{n+1}
   + \cdots \,.
$$
Using this and extracting the terms of order $\hbar^{(n+1)}$ from the
equations satisfied by $\overline{\Gamma}^{[n]}$, namely $(4.64)-(4.66)$,
one finds
\begin{eqnarray}
&& \Lambda_{op} \overline{\Gamma}^{[n]}_{n+1} = 0 \,,\\
&& \frac{\delta \overline{\Gamma}^{[n]}_{n+1}}{\delta \lambda_a(x)}
    = 0 \,,\ \ \ \ \
  \frac{\delta \overline{\Gamma}^{[n]}_{n+1}}{\delta \lambda_y(x)} = 0 \,,
\\
&& \Sigma_a(x) \overline{\Gamma}^{[n]}_{n+1} = 0 \,, \ \ \ \ \ \
  \Sigma_y(x) \overline{\Gamma}^{[n]}_{n+1} = 0 \,,
\end{eqnarray}
Let $\overline{\Gamma}^{[n]}_{n+1,div}$ stand for the pole part of
$\overline{\Gamma}^{[n]}_{n+1}$. By repeating the steps going from $(4.33)$
to $(4.41)$, one can arrive at 
\begin{eqnarray}
\overline{\Gamma}^{[n]}_{n+1,div}
 =\, \alpha_1^{(n+1)} T_{(1)} + \alpha_2^{(n+1)} T_{(2)} 
    + \alpha_3^{(n+1)} T_{(3)} + \alpha_4^{(n+1)} T_{(4)}
    + \alpha_5^{(n+1)} T_{(5)} \,,
\end{eqnarray}
where $\alpha_1^{(n+1)}, \cdots, \alpha_5^{(n+1)}$ are constants of order
$(\hbar)^{n+1}$. Therefore, in order to cancel the  $n+1$ loop divergence
the counterterm of order $\hbar^{n+1}$ should be chosen as
\begin{eqnarray}
 \delta I^{[n+1]}_{count} = - \overline{\Gamma}^{[n]}_{n+1,div}
                      [\psi,\overline{\psi},W,C,\overline{C}] \,.
\end{eqnarray}
Adding this counterterm, the mass term and the $\lambda$ terms to
$\overline{I}_{{\rm eff}}^{[n]}$, one can express the effective action of
 order $\hbar^{n+1}$ as
\begin{eqnarray}
&& I_{{\rm eff}}^{[n+1]}
  [\psi,\overline{\psi},
   W,C,\overline{C},\lambda,K,L,n,l,p,n',l',p', g,g_1, M] \nonumber\\
&&\ \ \ \ \ \ \
 = I_{{\rm eff}}^{[0]}[\psi^{[0]},\overline{\psi}^{[0]},W^{[0]},C^{[0]},
    \overline{C}^{[0]},\lambda^{[0]},K^{[0]}, L^{[0]},n^{[0]},n^{'[0]},
   \cdots, g^{[0]}, g_1^{[0]}, M^{[0]}] \,,
\end{eqnarray}
where the bare fields and the bare parameters (to order $(\hbar)^{n+1}$) are
defined as
\begin{eqnarray}
&& W^{[0]}_{a \mu} = (Z_3^{[n+1]})^{1/2} W_{a \mu}
        =\big((Z_3^{[n]})^{1/2} - \alpha_1^{(n+1)}\big) W_{a \mu}\,,\ \ \ \
  L^{[0]}_a = (Z_3^{[n+1]})^{1/2} L_a  \,, \\
&&  W^{[0]}_{y \mu} = (Z_3^{'[n+1]})^{1/2} W_{y \mu}
        = \big((Z_3^{'[n]})^{1/2} - \alpha_2^{(n+1)}\big) W_{y \mu}\,,\\
&& C^{[0]}_a = (\widetilde{Z}_3^{[n+1]})^{1/2}C_a
     = \big(( \widetilde{Z}_3^{[n]})^{1/2}
           + (-\alpha_3^{(n+1)} + \alpha_1^{(n+1)}) \big) C_a \,,\\
&& \overline{C}^{[0]}_a = (\widetilde{Z}_3^{[n+1]})^{1/2} \overline{C}_a \,,
 \ \ \ \ \
 K^{a[0]}_{\mu} = (\widetilde{Z}_3^{[n+1]})^{1/2} K^a_{\mu} \,,\\
&& C^{[0]}_y = (\widetilde{Z}_3^{'[n+1]})^{1/2}C_y
   = \big((\widetilde{Z}_3^{'[n]})^{1/2}
          + (-\alpha_3^{(n+1)} + \alpha_2^{(n+1)}) \big) C_y \,,\\
&& \overline{C}^{[0]}_y = (\widetilde{Z}_3^{'[n+1]})^{1/2} \overline{C}_y \,,
  \ \ \ \
  K^{y[0]}_{\mu} = (\widetilde{Z}_3^{[n+1]})^{1/2} K^y_{\mu}\,,\\
&&  \nu_L^{[0]} = (Z_{\nu L}^{[n+1]})^{1/2} \nu_L
     = \big((Z_{\nu L}^{[n]})^{1/2}
            - \alpha_4^{(n+1)} \big) \nu_L
   \,,\ \ \ \
  \overline{\nu}_L^{[0]}
    = (Z_{\nu L}^{[n+1]})^{1/2} \overline{\nu}_L \,,\\
&& e_L^{[0]} = (Z_{eL}^{[n+1]})^{1/2} e_L
         = (Z_{\nu L}^{[n+1]})^{1/2} e_L \,,\ \ \ \
   \overline{e}_L^{[0]}
                 = (Z_{eL}^{[n+1]})^{1/2} \overline{e}_L \,,\\
&& e_R^{[0]} = (Z_{eR}^{[n+1]})^{1/2} e_R
        = \big((Z_{eR}^{[n]})^{1/2}
               - \alpha_5^{(n+1)}\big) e_R \,,\ \ \ \
  \overline{e}_R^{[0]}
                = (Z_{eR}^{[n+1]})^{1/2} \overline{e}_R  \,,\\
&&  n^{[0]} = (Z_{(n)}^{[n+1]})^{1/2} n
      = \big((Z_{(n)}^{[n]})^{1/2}
             + (-\alpha_3^{(n+1)} + \alpha_4^{(n+1)}) \big) n
      \,,\ \ \ \
    n^{'[0]} = (Z_{(n)}^{[n+1]})^{1/2} n' \,,\\
&& l^{[0]}  = (Z_{(l)}^{[n+1]})^{1/2} l
           = (Z_{(n)}^{[n+1]})^{1/2} l  \,, \ \ \ \
   l^{'[0]} = (Z_{(l)}^{[n+1]})^{1/2} l' \,,\\
&&  p^{[0]}  = (Z_{(p)}^{[n+1]})^{1/2} p
     = \big( (Z_{(p)}^{[n]})^{1/2} - \alpha_3^{(n+1)}
                   + \alpha_5^{(n+1)} \big) p \,,\ \ \ \
   p^{'[0]} = (Z_{(p)}^{[n+1]})^{1/2} p'  \,,\\
&&  g^{[0]} = Z_g^{[n+1]} g = (Z_3^{[n+1]})^{-1/2} g \,,\ \ \ \
   g_1^{[0]} = Z_g^{'[n+1]} g_1 = (Z_3^{'[n+1]})^{-1/2} g_1 \,,\\
&&  g^{[0]} = Z_g^{[n+1]} g = (Z_3^{[n+1]})^{-1/2} g
     \,,\ \ \ \
   g_1^{[0]} = Z_g^{'[n+1]} g_1  = (Z_3^{'[n+1]})^{-1/2} g_1 \,,\\
&&  M^{[0]} = Z_M^{[n+1]} M = (Z_3^{[n+1]})^{-1/2} M \,,
\end{eqnarray}
and $ \lambda_a^{[0]}, \lambda_y^{[0]} $ are
\begin{eqnarray}
 \lambda_a^{[0]} = (Z_3^{[n+1]})^{-1/2} \lambda_a \,, \ \ \ \
  \lambda_y^{[0]} = (Z_3^{'[n+1]})^{-1/2} \lambda_y \,.
\end{eqnarray}
Therefore, in terms of such bare fields and bare parameters,
 $ \overline{\Gamma}^{[n+1]}$ can be expressed as
\begin{eqnarray}
  \overline{\Gamma}^{[n+1]}
  && [W,C,\overline{C},
  \psi,\overline{\psi},K,L,n,l,p,n',l',p', g,g_1, M] \nonumber\\
 && = \overline{\Gamma}^{[0]}[W^{[0]},C^{[0]},\overline{C}^{[0]},
   \psi^{[0]},\overline{\psi}^{[0]},K^{[0]}, L^{[0]},n^{[0]},n^{'[0]},
   \cdots, g^{[0]}, g_1^{[0]}, M^{[0]}] \,.
\end{eqnarray}
From this one can conclude that $\overline{\Gamma}^{[n+1]}$, under
$(4.23)$, satisfies $(4.64)-$$(4.66)$ and is finite to order $\hbar^{n+1}$.
Since the theory can be renormalized to one loop the renormalisability has
been proven.
\par
\vspace{8mm}
\begin{center}
{\bf V}.\ \ Concluding Remarks 
\end{center} \par
\ \par
     By taking into account the original constraint conditions and the
additional condition  we have carried out the quantization  of the
 SU$_L$(2) $\times$ U$_Y$(1) electroweak theory with the W Z mass term and
construct the ghost action in a way similar to that used for the massive
SU(n) theory [1]. We have also shown that when the $\delta-$ functions
appearing in the path integral of the Green functions and representing the
constraint conditions are rewritten as Fourier integrals with Lagrange
multipliers $\lambda_a$ and $\lambda_y$, the total effective action
consisting of the Lagrange multipliers, ghost fields and the original fields
is BRST invariant. Furthermore, by comparing with the massless theory and
with the massive SU(n) theory we have found the general form of the divergent
part of the generating functional for the regular vertex functions and proven
the renormalisability of the theory. It has also been clarified that the
renormalisability of the theory with the  W Z mass term is ensured by the
renormalisability of the massless theory and the massive  SU(n) theory. \par
    If the harmlessness of the W Z mass term had been proven at the begining
of 1960s, the SU$_L$(2) $\times$ U$_Y$(1) electroweak theory without the
Higgs mechanism would have been deeply studied and tested. Today, the
standard model of the electroweak theory has achieved great successes and
the whereabouts of the Higgs Bosons is still unknown. It is therefore
reasonable to ask if such successes really depends on the Higgs mechanism
and to pay attention to the theory without the Higgs mechanism.
\par
\vspace{2mm}
\begin{center}
\bf{ACKNOWLEDGMENTS}
\end{center} \par
    We are grateful to Professor Yang Li-ming for helpful discussions. This
work was supported by National Natural Science Foundation of China and
supported in part by Doctoral Programm Foundation of the Institution of
Higher Education of China.
\vspace{4mm}
\par
\ \par
\begin{center}
{\large \bf Refernces}
\end{center}
\par  \noindent
[1]\ Ze-Sen Yang, Zhining Zhou, Yushu Zhong and Xianhui Li,
        hep-th/9912046 7 Dec 1999.
\par  \noindent
[2]\ Ze-Sen Yang, Xianhui Li, Zhining Zhou and Yushu Zhong,
        hep-th/9912034 5 Dec 1999.
\par  \noindent
[3]\ M.Carena and C.Wagner, Phys. Rev. {\bf D37}, 560(1988).
\par  \noindent
[4]\ R.Delbourgo and G.Thompson, Phys. Rev. Lett. {\bf 57}, 2610(1986).
\par  \noindent
[5]\ M.Carena and C.Wagner, Phys. Rev. {\bf D37}, 560(1988).
\par  \noindent
[6]\ R.Delbourgo and G.Thompson, Phys. Rev. LeTT. {\bf 57}, 2610(1986).
\par  \noindent
[7]\ A.Burnel, Phys. Rev. {\bf D33}, 2981(1986);{\bf D33}, 2985(1986);.
\par  \noindent
[8]\ T.Fukuda, M.Monoa, M.Takeda and K.Yokoyama, \par\ \ \ \ 
Prog. Theor. Phys. {\bf 66},1827(1981);{\bf 67},1206(1982);{\bf 70},284(1983).
\par  \noindent
[9]\ S.L.Glashow, Nucl. Phys. {\bf 22}, 579(1961).
\par  \noindent
[10]\ Ze-Sen Yang, Xianhui Li, Zhining Zhou and Yushu Zhong,
       Manusript, Mar (2000).
\par  \noindent
[11]\ L.D. Faddeev and V.N. Popov, Phys. Lett. {\bf B25}, 29(1967).
\par  \noindent
[12]\ B.S. De Witt, Phys. Rev. {\bf 162},1195,1239(1967).
\par  \noindent
[13]\ L.D. Faddev and A.A.Slavnov, Gauge Field: Introduction to Quatum Theory,
\par\ \ \ \  The Benjamin Cummings Publishing Company, 1980.
\par  \noindent
[14]\ G.H.Lee and J.H.Yee, Phys. Rev. {\bf D46}, 865(1992). 
\par  \noindent
[15]\ C.Itzykson and F-B.Zuber, Quantum Field Theory, McGraw-Hill,
      New York, 1980.
\par  \noindent
[16]\ Yang Ze-sen, Advanced Quantum Machanics, Peking University Press,
     2-ed.  Beijing, 1995. 
\par
\end{document}